\documentclass{aa}
\usepackage{afterpage}
\usepackage{graphicx}
\usepackage{subcaption}
\usepackage{newtxtext}
\usepackage[frenchmath,varg]{newtxmath}
\usepackage{xcolor}
\usepackage[colorlinks=true,linkcolor=bluea,allcolors=blue]{hyperref}

\newcommand{\pc}{\ensuremath{\,\mathrm{pc}}}
\newcommand{\kpc}{\ensuremath{\,\mathrm{kpc}}}
\newcommand{\Mpc}{\,\mathrm{Mpc}}
\newcommand{\GHz}{\,\mathrm{GHz}}
\newcommand{\MHz}{\,\mathrm{MHz}}

\newcommand{\Myr}{\,\mathrm{Myrs}}
\newcommand{\kyr}{\,\mathrm{kyrs}}

\newcommand{\Jy}{\,\mathrm{Jy}}
\newcommand{\jykms}{\,\mathrm{Jy\,km s^{-1}}}
\newcommand{\kkmspc}{\,\mathrm{K\,km s^{-1}\,pc^2}}
\newcommand{\ang}{\,\text{\normalfont\AA}}
\newcommand{\K}{\,\mathrm{K}}

\newcommand{\msun}{\,\mathrm{M_\odot}}

\newcommand{\kms}{\,\mathrm{km\,s^{-1}}}
\newcommand{\ergs}{\,\mathrm{erg\,s^{-1}}}
\newcommand{\coto}{\ensuremath{\mathrm{CO(2-1)}}}
\newcommand{\cott}{\ensuremath{\mathrm{CO(3-2)}}}
\newcommand{\cooz}{\ensuremath{\mathrm{CO(1-0)}}}
\newcommand{\hone}{\ensuremath{\mathrm{HI}}}
\newcommand{\htwo}{\ensuremath{\mathrm{H_2}}}
\newcommand{\oiii}{\ensuremath{\mathrm{[OIII]}}}
\newcommand{\xco}{\ensuremath{\mathrm{X_{CO}}}}
\renewcommand\deg{^{\circ}}

\begin{document}

\titlerunning{Dynamically unsettled gas in NGC6328}
\authorrunning{M. Papachristou et al.}

\title{A plausible link between dynamically unsettled molecular gas and the radio jet in NGC 6328}

\subtitle{}

\author{M. Papachristou\inst{1,2}, K.M. Dasyra\inst{2,1}, J.A. Fernández-Ontiveros\inst{1,3,4}, A. Audibert\inst{1,5,6}, I. Ruffa\inst{7,1}, F. Combes\inst{8}, M. Polkas\inst{1}, A. Gkogkou\inst{9}}

\institute{
     National Observatory of Athens (NOA), Institute for Astronomy, Astrophysics, Space Applications and Remote Sensing (IAASARS), GR–15236, Greece
     \and
     Department of Astrophysics, Astronomy \& Mechanics, Faculty of Physics, National and Kapodistrian University of Athens, Panepistimiopolis Zografou, 15784, Greece
     \and
     Istituto di Astrofisica e Planetologia Spaziali (INAF–IAPS), Via Fosso del Cavaliere 100, I--00133 Roma, Italy
     \and
     Centro de Estudios de F\'isica del Cosmos de Arag\'on (CEFCA), Plaza San Juan 1, E--44001, Teruel, Spain
     \and
     Instituto de Astrof\' isica de Canarias, Calle V\' ia L\'actea, s/n, E-38205, La Laguna, Tenerife, Spain  
     \and
     Departamento de Astrof\' isica, Universidad de La Laguna, E-38206, La Laguna, Tenerife, Spain
     \and
     Cardiff Hub for Astrophysics Research \&\ Technology, School of Physics \&\ Astronomy, Cardiff University, Queens Buildings, The Parade, Cardiff, CF24 3AA, UK
     \and
     Observatoire de Paris, LERMA, Coll\`ege de France, CNRS, PSL Univ., Sorbonne Univ., F--75014 Paris, France
     \and
     Aix Marseille Univ., CNRS, CNES, LAM, Marseille, France
      }

\date{}

\abstract{We report the detection of outflowing molecular gas at the center of the nearby radio galaxy NGC6328 (z=0.014), which has a gigahertz-peaked spectrum radio core and a compact ($2\pc$) young double radio lobe tracing jet. Utilizing Atacama Large Millimeter/submillimeter Array (ALMA)  $\cott$ and $\coto$ observations, as well as a novel code developed to fit the 3D gas distribution and kinematics, to study the molecular gas kinematics, we find that the bulk of the gas is situated within a highly warped disk structure, most likely the result of a past merger event. Our analysis further uncovers, within the inner regions of the gas distribution ($R<300\pc$) and at a position angle aligning with that of the radio jet ($150\deg$), the existence of two anti-diametric molecular gas structures kinematically detached from the main disk. These structures most likely trace a jet-induced cold gas outflow with a total lower limit mass of $2\times 10^6\msun$ mass, corresponding to an outflow rate of $2\,\mathrm{M_\odot\,yr^{-1}}$ and a kinetic power of $2.7\times 10^{40}\,\mathrm{erg\,s^{-1}}$. The energy required to maintain such a molecular outflow is aligned with the mechanical power of the jet.}


\maketitle

\section{Introduction}
\label{sec:introduction}

A central question in the study of galaxy evolution is the role and importance of the central engines of active galaxies, active galactic nuclei (AGN), in regulating star formation (SF). As a supermassive black hole (SMBH) grows in the center of galaxies, due to the accretion of gas \citep{antonucci1993}, it releases energy in the form of radiation pressure or in the form of radio jets (i.e., collimated plasma that is often relativistic). This energy can be coupled to the surrounding interstellar medium (ISM), affecting the molecular gas that is essential for SF \citep{krumholz2009}. Thus, AGN feedback can suppress gas accretion and quench SF, ultimately shaping the evolution of the host galaxy \citep{croton2006a, sijacki2007a, harrison2018}.

\afterpage{
\begin{figure*}[h]
    \begin{center}
        \includegraphics[width=17.5cm]{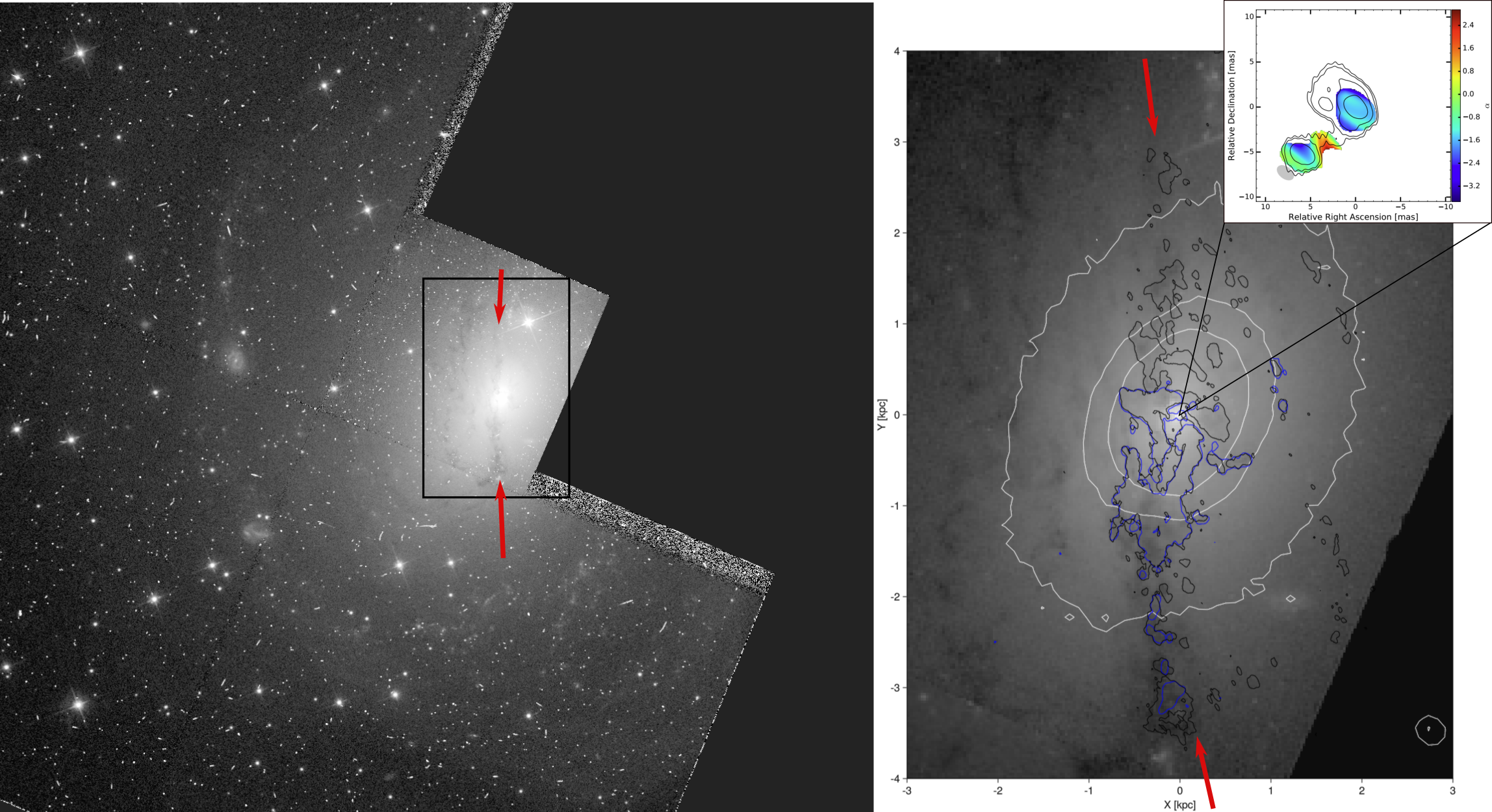}
        \caption{Hubble Space Telescope (WFPC2) view of NGC6328. Left: Grand-scale ($\sim16\kpc$) image (obtained from \href{https://commons.wikimedia.org/wiki/File:NGC_6328_hst_05479_606.png}{Wikimedia}). Right: Zoomed-in view of the central $4\kpc$ (indicated also in the black box on the left). \coto\ (black) and \cott\ (blue) $4\sigma$ contours of the maximum flux per channel. The central north-south oriented dust lane, correlated with the CO emission, is indicated with red arrows, along with a less prominent feature to the west marking the start of a faint spiral arm. The elliptical contours represent the stellar distribution from VLT in the J band. In the top right inset, we include the bipolar radio map from \cite{angioni2019} and the spectral index between $8.4\GHz$ and $22.3\GHz$.}
        \label{fig:hst}
    \end{center}
\end{figure*}
}

The role of relativistic jets in AGN feedback has primarily been recognized through their ability to heat the circumgalactic environment and prevent gas cooling and inflows by inflating thermal bubbles (so-called "maintenance mode" feedback) (e.g., \citealt{fabian2012}; \citealt{mcnamara2012}). Alternatively, simulations have shown that jets can also be highly efficient at coupling their power to the local dense medium as they propagate through the ISM, providing a source of shocks and ram pressure and inducing turbulence (\citealt{wagner2011}; \citealt{wagner2012}; \citealt{mukherjee2016}; \citealt{mukherjee2018}). When the clouds are dissipated to very hot and tenuous conditions, or even accelerated to velocities greater than the escape velocity of the gravitational potential, SF declines as the galaxy is starved of its star-forming fuel. This has been observed in simulations in the aforementioned studies, although only a small fraction of the ISM can reach such high velocities and most of the gas will eventually fall back in a galactic fountain. Also, according to the simulations, the total impact a jet will have on the surrounding gas depends on numerous parameters, mainly its power and orientation relative to the gas disk and the clumpiness of the disk. Powerful jets ($\sim 10^{45}\ergs$) tend to drill easily through the ISM and escape to intergalactic space much faster than low-power jets ($\sim 10^{43}\ergs$). Low-power jets get diffused in the dense gas environment for larger timescales. As a result they tend to power high-pressure bubbles, greatly affecting the ISM \citep{talbot2021,mukherjee2016}. 

Recent studies have shown that compact radio sources, associated with low-powered parsec-scale young radio jets are quite common \citep{baldi2018,baldi2021}. Young jets have been also associated with peaked-spectrum (PS) radio sources \citep{odea2021} which exhibit a characteristic turnover in their radio spectral energy distribution (SED) due to their interaction with the ISM. 
There are two proposed mechanisms to which the observed peak can be attributed, namely the free-free absorption (FFA) of the synchrotron radiation by the ionized, shocked, and dense ISM \citep{bicknell1997} and synchrotron self-absorption (SSA) \citep{Kellermann1969}. Radio spectra obtained from simulations of young jet-ISM interactions \citep{bicknell2018} reproduce the observed turnover frequency and radio source size correlation. This makes PS sources suitable candidates for understanding the impact of young jets in the ISM and their significance in AGN feedback. 

\begin{figure*}[h]
    \begin{center}
        \includegraphics[width=16cm]{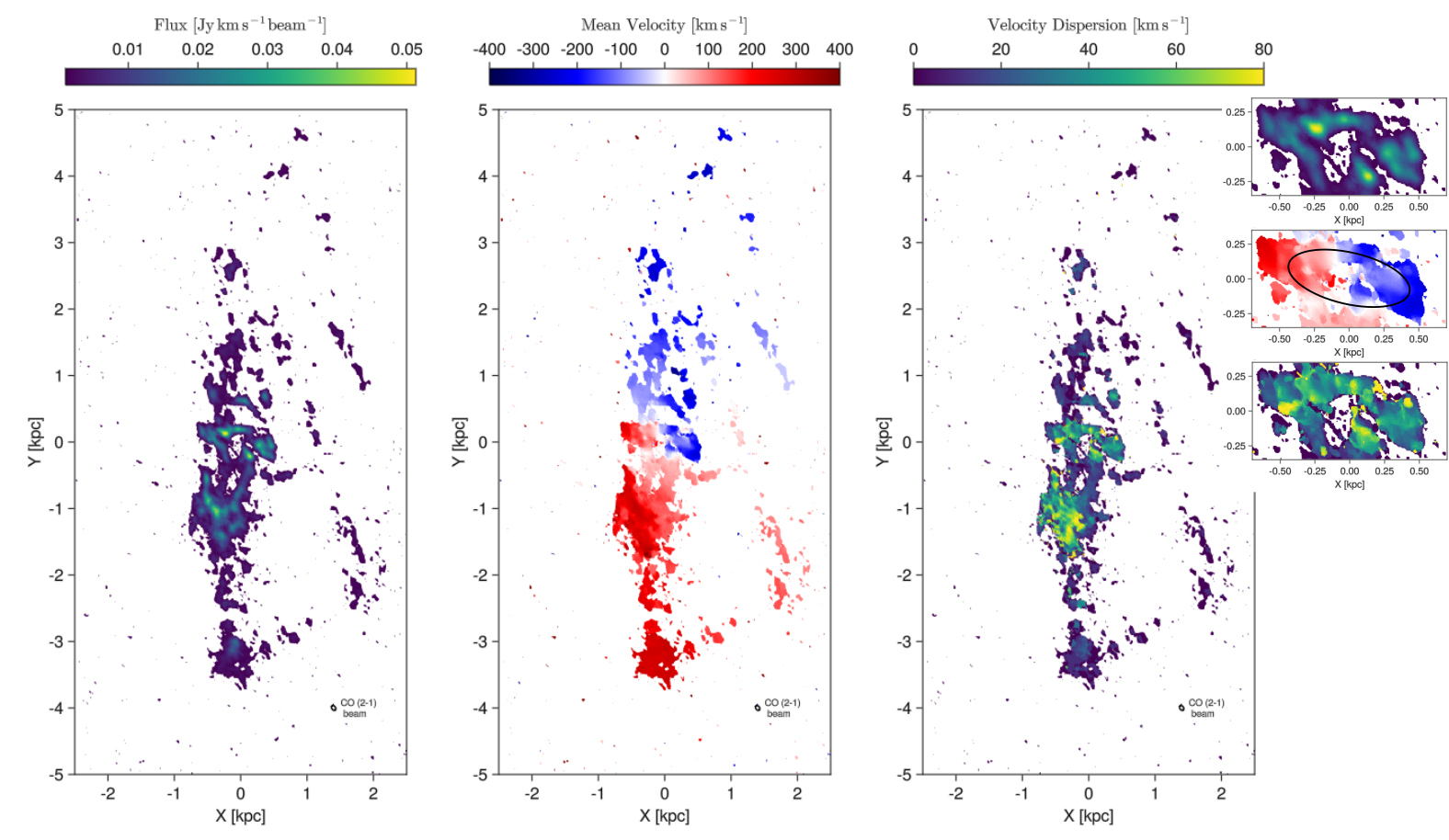}
        \caption{From left to right: Zeroth, first, and second moments of the \coto\ emission. The ellipse in the lower right corner of each panel represents the observational beam. The inset on the right provides a zoomed-in view of the central inner disk-like structure, also highlighted by a black ellipse.}
        \label{fig:CO21moments}
    \end{center}
\end{figure*}

NGC 6328 is an early-type galaxy with a well-studied gigahertz-peaked source (GPS) radio core (PKS B1718-649) that exhibits clear evidence of out-of-dynamic-equilibrium kinematics in both its cold and warm gas phases \citep{maccagni2014, maccagni2016a, maccagni2018}. Due to its proximity ($D=62.88\Mpc$ at $z=0.014$), NGC6328 is an ideal target for studying the impact of young jets on the ISM with the Atacama Large Millimeter/submillimeter Array (ALMA) and it is part of the CO-ARC survey on the molecular gas content of radio galaxies \citep{audibert2022}.

PKS B1718-649 shows a typical GPS spectrum, with a turnover point at around $3.2\GHz$  \citep{tingay1997}. The radio spectrum is best modeled with the FFA model \citep{tingay1997}, while close monitoring of the source reveals significant variability, likely due to a mixture of intrinsic processes associated with the dense local environment and adiabatic losses of the synchrotron-emitting lobes \citep{tingay2015}. High-resolution ($\sim 0.3\pc$) very long baseline interferometry (VLBI) observations resolve the source into a bipolar structure, with the two lobes being separated by a projected distance of $\sim 2\pc$ \citep{tingay1997, tingay2003a}. The spectral index (see Figure~\ref{fig:hst} inset) between $8.4\GHz$ and $22.3\GHz$ suggests that the core of the young radio source is strongly absorbed in these frequencies, and is located between the two lobes. The structure is unresolved in lower-resolution ($29.7\farcs$ or $8.7\kpc$) observations at $1.4\GHz$ from the Australia Telescope Compact Array (ATCA), in which the total measured flux ($3.98\Jy$) is comparable with the VLBI emission \citep{maccagni2014}. 

Particularly interesting is the warped and complex kinematics and structure of the cold gas disk \citep{maccagni2014}, which is indicative of an external accretion event from an initially misaligned disk \citep[see e.g.,][]{vandevoort2015,ruffa2019}. This event could provide valuable insights into the feeding mechanism of the AGN and the transport of angular momentum in galaxy disks. This complex structure is also visible in the dust distribution seen in Hubble Space Telescope (HST) observations through the Wide Field and Planetary Camera 2 (WFPC2), where it appears as a near edge-on north-south oriented dust lane in the central $23\arcsec$ of the bulge. At larger distances, faint spiral features of dust and stars are also evident, resembling a more face-on disk orientation. This is prominent in the Neil Gehrels Swift Observatory maps obtained using Ultraviolet/Optical Telescope (UVOT) filters (UVM2 at $2246\ang$ and UVW2 at $1928\ang$). Overall, the optical surface brightness of NGC6328 is dominated by a sersic 1/4 elliptical bulge with a position angle (PA) of $145\deg$ and an axis ratio (b/a) of $0.72$ \citep{veron-cetty1995}.

Observations of atomic hydrogen in NGC 6328 by \citealt{maccagni2014} revealed a massive \hone\ disk with a total mass of $M_{\hone} = 1.1\times10^{10}\msun$. Similar to the molecular gas, the kinematics of the \hone\ disk can be modeled as a warped disk, with an almost constant circular velocity of $220\kms$.  In the central $30\arcsec$ (or $8.5\kpc$ for the adopted Lambda cold dark matter ($\Lambda$CDM) cosmology with $H_0=69.32\kms \Mpc^{-1}$, $\Omega_m=0.29$, $\Omega_\Lambda=0.71$), the disk is highly inclined, with an inclination angle, $\mathrm{i}$, of $\sim 90\deg$, and oriented along the PA of the observed dust lane ($180\deg$). In outer regions, up to $80\arcsec$ (or $23\kpc$), the disk slowly warps to a face-on geometry ($\mathrm{i}=30\deg$ and $\mathrm{PA}=110\deg$), consistent with the galaxy's spiral features. Further out (past the 80$\arcsec$ radius), asymmetric features are present northwest and south of the disk. These deviate from regular rotation and are attributed to an ongoing merger or interaction (\cite{veron-cetty1995}, \cite{maccagni2014}

Very Large Telescope (VLT) Spectrograph for INtegral Field Observations in the Near Infrared (SINFONI) observations of warm molecular hydrogen at $2.12\,\mu m$ (1-0) S(1) and $1.95\,\mu m$ (1-0) S(3) by \cite{maccagni2016} showed \htwo\ emission within the central $700\pc$ of the galaxy following a nearly edge-on inner disk at a PA of $85\deg$, almost perpendicular to the dust lane. The \htwo\, at an excitation temperature of $1100-1600 \K$ has an estimated mass of $10^4 \msun$. In the outer disk, two  warm \htwo\ blobs have been detected  in opposite directions, at a projected distance of $1-1.5 \kpc$, close to the radio jet axis. This axis happens to be nearly parallel to the major axis of the stellar isophote curves ($\mathrm{PA}\sim 150\deg$).

In this paper, we present a new kinematic analysis of the molecular gas in NGC 6328, utilizing previous CO(2-1) observations \citep{maccagni2018} and more recent CO(3-2) data obtained with ALMA.
The paper is structured as follows: Section \ref{sec:ALMAdata} provides a description of the ALMA observations. In Section \ref{sec:modeling}, we present our novel approach to the kinematic modeling of the cold gas disk of the galaxy and the optimal 3D model that describes its structure. There, emphasis is given to the nature of the clouds that strongly deviate from the optimal model in the form of outflowing gas. Finally, in Section \ref{sec:discussion}, the possible connections of the outflows with the AGN and the jet are discussed.

\begin{figure*}[h]
    \begin{center}
        \includegraphics[width=16cm]{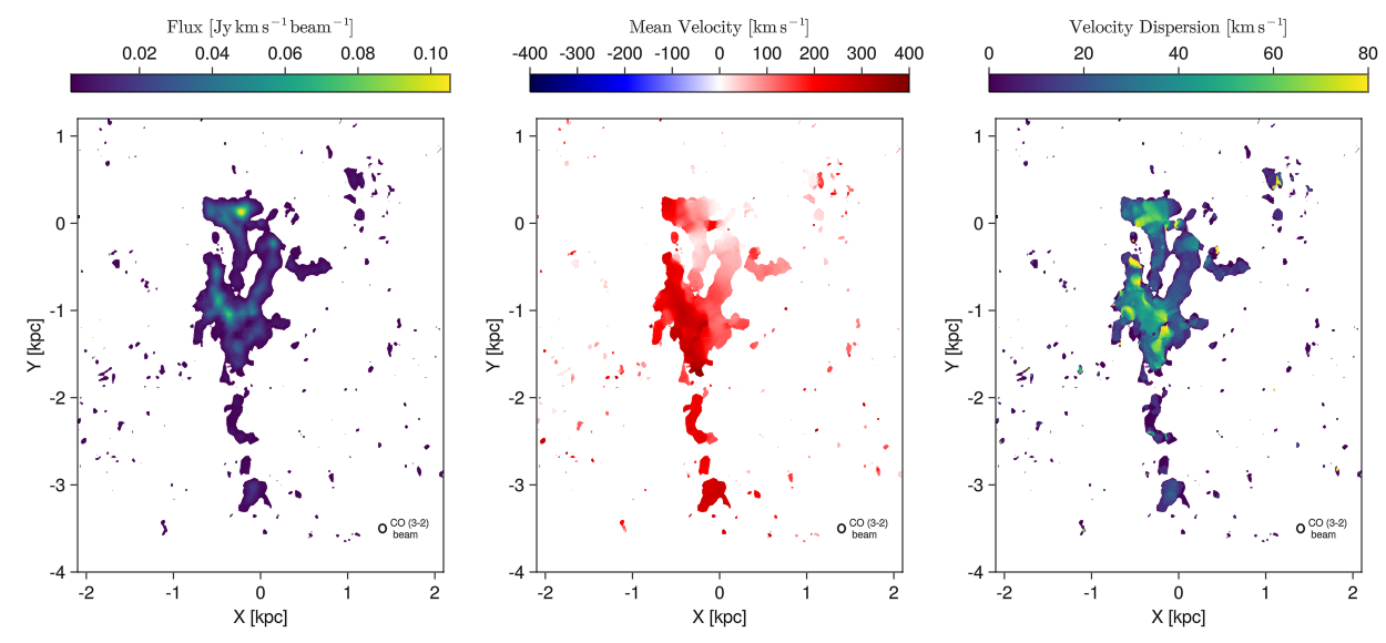}
        \caption{From left to right: Zeroth, first, and second moments of the \cott\ emission. The black ellipse represents the beam of the observation. Due to spectral coverage limitation, only the positive velocities are observed, and thus only the redshifted part of the disk is visible.}
        \label{fig:CO32moments}
    \end{center}
\end{figure*}
    
\section{Observations}
\subsection{ALMA data reduction}
\label{sec:ALMAdata}

NGC6328 was observed using ALMA Band 6 at a rest frame frequency, $\nu_\mathrm{rest}$, of $230.5\GHz$ (project ID: \#2015.1.01359.S, PI: F. Maccagni) in September 2016, and using Band 7 at $\nu_\mathrm{rest}=345.8\GHz$ (project ID: \#2017.1.01638.S, PI: S. Kameno) in January 2018. The phase center of the observations was the position of the nucleus ($\delta_\mathrm{RA}=$ 17:23:41.03 and $\delta_\mathrm{DEC}=$-65:00:36.615), with a single pointing covering a field of view (FoV) of 25\arcsec\ in Band 6 and 18\arcsec\ in Band 7.

The ALMA Band 6 observations were carried out with 39 12-m antennas and an extended array configuration with baseline lengths ranging from 15 to 2483\,m. The on-source integration time was 1.85 hours. The correlator setup was selected to center the CO(2-1) line with a velocity range of $\pm$1200\, km/s in the 1.86\, GHz bandwidth and a channel width of $7.8\MHz$ ($20\kms$). 

The data were calibrated using the Common Astronomy Software Applications package (\textsc{CASA}) \citep{mcmullin}, version 4.7, and the calibration scripts provided by the ALMA archive. J1617-5848, J2056-4714, and J1703-6212 were respectively used as flux, bandpass, and phase calibrators, resulting in 15\% accuracy. The data were imaged using the CASA {\tt tclean} task with Briggs weighting and a robust parameter of two, in order to maximize the sensitivity, resulting in a synthesized beam of $0\farcs 27 \times 0\farcs 19$ at PA=$33.75\deg$. The images were binned to a $20\kms$ ($15.2\MHz$) spectral resolution and corrected for primary beam attenuation.

The continuum map was made using natural weighting, resulting in a signal-to-noise ratio (SNR) of 400. After five iterations of phase-only self-calibration and one iteration amplitude and phase self-calibration, the SNR was raised to 7912 with an rms of $0.04\,\mathrm{mJy/beam}$. The restoring beam is 0\farcs194$\times$0\farcs107 at $\mathrm{PA}=21.5\deg$. In the continuum map the radio source is unresolved. 
The \coto\ line emission cube was obtained after subtraction of the continuum in the line-free channels with the \textsc{CASA} task \texttt{uvcontsub}. The rms noise was estimated at 0.04\,mJy/beam, with a restoring beam of $0\farcs 272 \times 0\farcs186$ at $\mathrm{PA}=34.8\deg$. 

The ALMA Band 7 observations were carried out with 44 antennas and baselines ranging from 15 to 1400m. The shortest baseline provides a larger recoverable angular scale of about $11\arcsec$. The total integration time was 44.7 minutes. The galaxy was observed in dual-polarization mode with $2\GHz$ total bandwidth and 128 channels per spectral window (SPW), corresponding to $\sim13.7\kms$ velocity resolution. The choice of the tuning setup was to provide simultaneous observations of \cott\,, HCO$\rm^{+}$(4-3), and CS(7-6) emission lines, and therefore the SPW of the \cott\ line emission did not center the observed frequency of the \cott\ line, missing out the negative velocities. The data reduction was performed using \textsc{CASA} version 5.4.0-70. For the phase and bandpass calibrations, J1703-6212 and J1647-6438 were used, respectively. The flux was calibrated using J1427-4206, assuming that at 350GHz the flux density is 2.19\, Jy (for a spectral index of -0.52), resulting in an accuracy of 8\%. 
The continuum map was made using the CASA {\tt tclean} task with natural weighting, resulting in a SNR of 186. After five iterations of phase-only self-calibration, and one iteration of amplitude and phase self-calibration, a SNR of 5063 was reached, with an rms of 0.044\,mJy/beam and a restoring beam of $0\farcs339\times0\farcs269$ at $\mathrm{PA}=6.9\deg$. After subtracting the continuum and cleaning using the same criteria as in Band 6, we obtained a \cott\ data cube with a restored beam of $0\farcs 320 \times 0\farcs275$ at $\mathrm{PA}=2.2\deg$, and an rms noise of 0.297\,mJy/beam per channel of $20\kms$.

\subsection{Ancillary data}
\label{sec:xshooter_data}
We also used archival ancillary data from the X-shooter instrument on the VLT, which provides simultaneous coverage of the optical and near-infrared spectral ranges. These data (PI: Maccagni and program name: 105.20QD.001) have not yet been published. They have only been publicly released and calibrated through the standard pipeline. We used the X-shooter data to present the basic ionized gas properties of NGC 6328, particularly in the central regions of the galaxy. While a detailed analysis of these data is beyond the scope of this paper, we note that they provide valuable complementary information to the ALMA data analysed here, and will be discussed in more detail in future works.

\section{A new, physically motivated 3D fitting Bayesian tool for the gas kinematics and distribution model}
\label{sec:modeling}

To investigate the kinematics of the molecular gas in NGC6328, we developed a novel Bayesian tilted-ring code that extracts the galaxy's intrinsic 3D structure from observational data. Tilted-ring models are common in kinematic studies, modeling galaxies either in two dimensions (through average velocity moment maps) or in three dimensions (observational cubes), with \textsuperscript{3D}{\sc Barolo} being a notable example \citep{diteodoro3DBaroloNew3D2015}. Traditional tilted-ring models, however, typically either treat each ring as an isolated entity or connect them using simple interpolation schemes, without accounting for the physical interactions between adjacent rings. Recognizing that the molecular gas in NGC6328 is likely part of a strongly warped structure, arising from a differential precession process due the presence of a nonspherical potential, our code incorporates additional physical principles to depict the galaxy's 3D structure and its temporal evolution in a more interconnected manner. 
This method provides a coherent connection between rings, which is especially crucial in the context of warped structures, and which helps mitigate parameter degeneracies related to overfitting. This is particularly beneficial in regions where distinct parts of the disk with different intrinsic rotation velocities overlap in projection. As a result, our approach enhances the reliability of inferences regarding the disk’s kinematics and geometry. 

Additionally, our code includes an alternative model that employs the traditional interpolation schemes for the rings' geometrical properties in the galaxy plane, akin to traditional methods. This alternative model is initialized using the resulting geometry derived from the physically motivated model, but allows for more flexibility in case the physical model might be overly constraining or not fully applicable. We start by describing this latter setup of our model.

\subsection{Generic tilted-ring model}
\label{sec:generic_model}
Our model starts by constructing the galaxy's disk as a series of concentric circular rings. The orientation of each ring within the galaxy's coordinate system, for a given radius, $r$, is determined by two angles: its polar angle, $\theta(r)$, and a modified azimuthal angle termed as the longitude of ascending node (LON), $\lambda(r)$. The LON specifies the point where the ring intersects the $x,y$ plane of the galaxy's coordinate system. Consequently, the normal vector to the plane of the ring, which is also indicative of the direction of the ring's angular momentum, is given by
\begin{equation}
 \hat{n}(r) = 
    \begin{pmatrix}
      -\sin \theta(r) \sin\lambda(r) \\
      \sin \theta(r) \cos\lambda(r) \\
      \cos \theta(r)
     \end{pmatrix}.
\end{equation}
{\noindent The} derived $\theta(r)$ and $\lambda(r)$ of each ring are then transformed into the observable inclination, $i(r)$, and PA, $\mathrm{PA}(r)$, using a reference “ring” for the main axis of the potential ($i_\mathrm{Bulge}$ and $\mathrm{PA}_\mathrm{Bulge}$; see Appendix~\ref{sec:Sky_coordinates_transformation}). 

For the rotation curve, $V_C(r)$, we used a gravitational potential comprised of three components: a SMBH, a bulge, and a dark matter halo. Further components (such as a molecular gas disk) can be provided to the code, but in the case of NGC6328 its mass is too small to significantly influence the kinematics. For the SMBH, we used a mass of $4.1\times10^8\msun$ \citep{willett2010}. For the dark matter halo component, we used the Navarro–Frenk–White profile \citep{navarroStructureColdDark1996}, with both the virial mass and the concentration parameter as free parameters (using the parameterization of \citealt{bullock2001}). For the bulge component, we utilized a Hernquist potential \citep{hernquist1990}. We assumed that noncircular velocities are negligible.

The best solution of the model is determined in a Bayesian framework. The model aims to maximize the likelihood that the predicted line-of-sight velocities ($V_{LOS}$) are consistent with the observed $V_{LOS}$ for all molecular gas clumps detected in the ALMA data at the same position in the sky.

To determine which observed clouds are reliably detected, we used a clump detection algorithm (see Appendix~\ref{sec:clump_detection}) for the ALMA cube. Then, given the set of $N$ observable molecular clouds with sky projected coordinates $\{(x_n,y_n,V_n)\}_{n=1}^N$, the code calculates for every cloud the minimum spatial distance (in sky coordinates) from every ring of our geometrical model. If this distance is below a critical limit (i.e., comparable with the beam size), the code calculates the projected velocity difference, $\delta V_{nk} = |V_n-V_\mathrm{los}(\phi _\mathrm{closest};V_{ck},i_k,\phi_{0k})|$, with $\phi _\mathrm{closest}$ corresponding to the closest spatial position on the projected ellipse of the $k$ ring.  When $\delta V$ is calculated for all the possible neighbouring rings, the code assigns the smallest one to the cloud, $m$, so $\delta V _n=\min{(\delta V_{nk})}$. 
Assuming a Gaussian likelihood of observing these velocity differences, we maximize the log-likelihood of
\begin{equation}
    \log L_\mathrm{ALMA}(\{\delta V_n\}^N_1|\bar{\theta})=-\frac{1}{2}\Big(\sum_{n=1}^{N} \frac{\delta V_n^2}{\sigma _V ^2}+N\log(2\pi\sigma _V^2)\Big),
\end{equation}
where $\sigma _V$ is a measure of the dispersion of the clouds' velocity from a perfectly rotating disk. 

For a more precise estimation of the large-scale kinematics we also incorporated HI rotational velocity data points from \cite{veron-cetty1995} and \cite{maccagni2014}, using them as a second likelihood distribution together with the ALMA data. We selected the HI data points only for the outer part of the galaxy in order to avoid any influence of the warp. This provides us with better accuracy on the dark matter component of the model, and thus a more accurate description of the overall velocity curve. 

The total posterior distribution was constructed using the joint likelihood distribution of the observable data with prior information regarding the stellar distribution and mass from \cite{maccagni2016a}, as well as the specific geometric constraints of the disk from the dust distribution and the shape of the HI disk mentioned in Section~\ref{sec:introduction}. The best set of parameters was estimated by sampling the posterior distribution using a Markov Chain Monte Carlo (MCMC) code \citep{affine-inv-mcmc-julia}.

\subsection{Physically connected ring model}
\label{sec:model_independent}

In the case of rings that are physically connected, the fitting procedure remains identical, with the exception that the geometrical properties of the disk, $\theta(t,r)$ and $\lambda(t,r)$, are time-dependent, and their evolution is governed by the precession induced by a nonspherical potential.

\begin{figure*}[h]
    \centering
    \includegraphics[width=15.5cm]{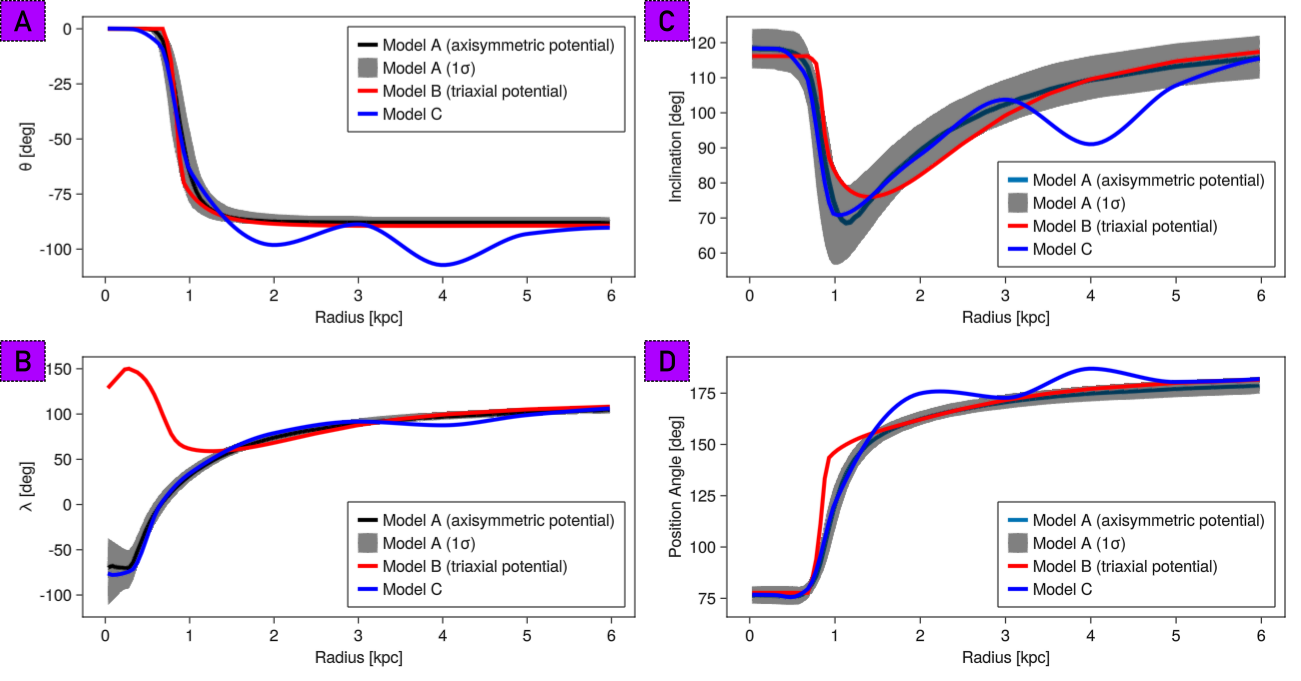}
    \caption{Modeling results: Main geometrical parameters as function of radius. The gray band represents the $1\sigma$ confidence level of the parameters' posterior distribution regarding the axisymmetric model. \textbf{A}: Azimuthal angle of the disk in the galaxy coordinates, \textbf{B}: Line of nodes in the galaxy coordinates, \textbf{C}: Inclination, \textbf{D}: PA.}
    \label{fig:geometry}
\end{figure*}

In a nonspherical potential, it is widely accepted that the gas will precess due to the torques induced by the gravitational potential of the stars, until it eventually relaxes back into alignment with one of the preferred axes, commencing either clockwise or counterclockwise rotation. Rings at smaller radii precess more rapidly due to their higher angular velocity, which leads to a warped disk structure. The differences in angular velocity and precession rates between neighboring rings generate internal torques. These internal torques serve as dissipative forces, promoting the alignment of each ring with its neighbors and eliminating the component of the gas's angular momentum that is perpendicular to the stellar angular momentum direction, and thus facilitating the settling of the gas to the potential's preferred axes. Over time, these processes result in the alignment or counter-alignment of the gas with the stellar angular momentum. This alignment occurs over timescales ranging from hundreds of millions to billions of years \citep{tohline1982,lake1983, steiman-cameron1988, sparke1996, davis2016}.

Assuming that precession is slow compared to the time taken for one complete orbit, the change in the orbital orientation (denoted as $\hat{n}(t,r)$) due to the asphericity of the galaxy and viscous forces can be calculated by averaging the torque induced over an orbital motion as follows:
\begin{align}
\label{eq:tilting}
\frac{d\theta}{dt} &=- \frac{T_\theta}{h \sin \theta} - f_\theta(t,r,\theta,\lambda)\\
\label{eq:twisting}
\frac{d \lambda}{dt} &=\frac{T_\lambda}{h \sin \theta} 
\end{align}
where $T_{\theta,\lambda}=-\frac{ \partial \langle\Phi\rangle (r,\theta,\lambda) }{ \partial (\lambda, \theta) }$ represents the torques exerted on the ring in the $\theta$ and $\lambda$ directions due to the asphericity of the potential, $\Phi$, and $h(r)=rV_c(r)$ is the orbital angular momentum per unit mass. The rotation curve, $V_c(r)$, is computed in the same manner as in the generic model, as described in the previous section.
The quantity, $\langle\Phi\rangle (r,\theta,\lambda)$, denotes the gravitational potential averaged along the circular orbit as $\langle\Phi\rangle (r,\theta,\lambda)=\frac{1}{2\pi}\oint_{r,\theta,\lambda}  \Phi\, d \phi$, where $0\le \phi \le 2\pi$.

The function, $f_{\theta}(t,r,\theta,\lambda)$, represents the effect of viscous forces arising from the interaction between adjacent rings.
For the evolution of the polar angle (tilting), as described in equation~\ref{eq:tilting}, we employ the dissipative force definition from \cite{steiman-cameron1988} to define the quantity
\begin{equation}
    f_{\theta}(t,r,\theta,\lambda) =\frac{3\sin2\theta}{2} \frac{t^2}{t^3_{e}}
\end{equation}
where $t_{e}$ is the timescale required for the gas to settle. 

This timescale is dependent on the geometric and physical characteristics of each ring and its neighbors. In their work, the authors approximated the inner torques between adjacent rings as resulting from viscous transport of angular momentum, through cloud-cloud collisions, computing $t_e$ via the gas viscosity and higher-order derivatives. 
In this study, we used the simple approach of \cite{tohline1982}, who defines $t_{e}=\frac{\alpha}{| \dot{\lambda}(r) - \dot{\lambda}_0 | }$, with the dot representing the time derivative. This method is based on the observation that the gas will have settled once it has precessed through an angle, $\alpha$, around the main axis. While the original work by \cite{tohline1982} used $\alpha=\pi/2$, we have chosen to treat this as a fitting parameter.

\begin{figure*}
    \begin{center}
        \includegraphics[width=15cm]{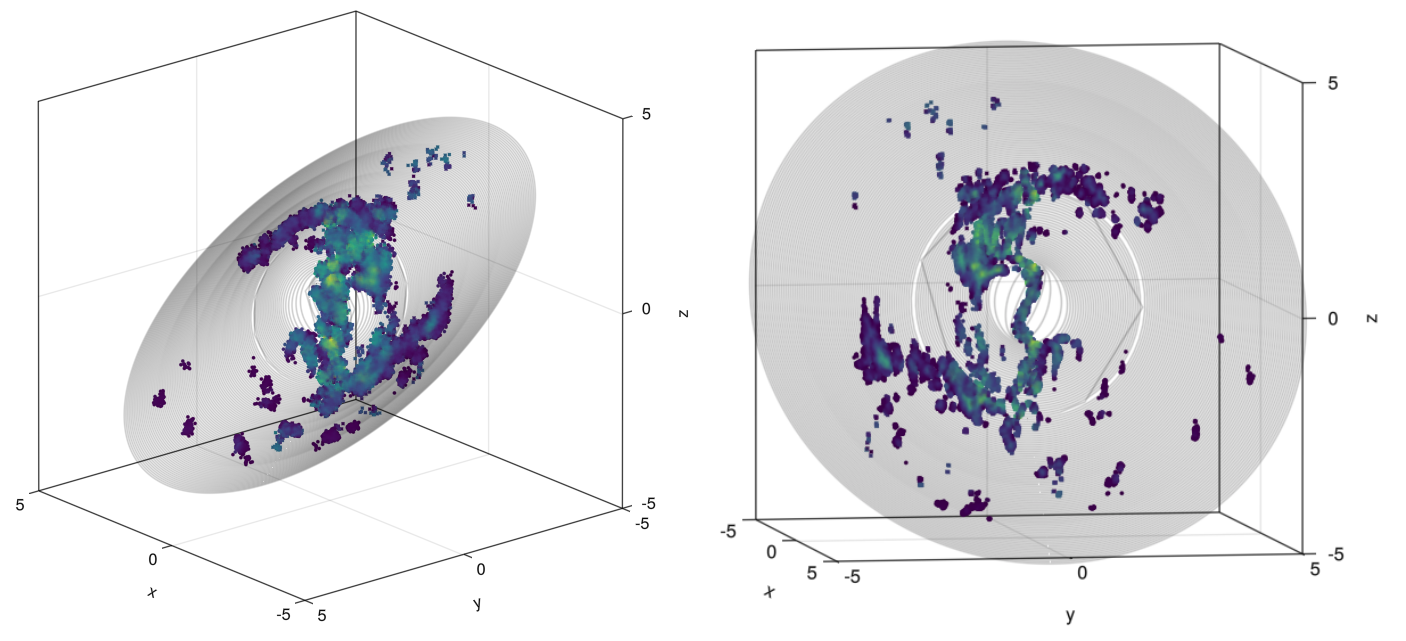}
    \caption{3D distribution of the \coto\ and \cott\ cloudlets and tilted ring of the model A at two random observable orientations. The X and Y axes represent the $\delta_{RA}$ and $\delta_{DEC}$ sky coordinates and the Z axis represents the line of sight. The color represents the integrated flux of each cloud. The axis units are in kpc. 
    }
    \label{fig:3d}
    \end{center}
\end{figure*}

Equation~\ref{eq:twisting} describes the precession of the disk (twisting). In the stress-free approximation, the effect of dissipative forces on twisting is considered negligible compared to their effect on tilting \citep{steiman-cameron1988}.

The potential that provides the necessary asymmetries for the above precession is assumed to be that of the bulge. Thus, we used a modified Hernquist bulge for the stellar density with total stellar mass $M$ and scale lengths $r_a, r_b,$ and $r_c$ (in case of an axisymmetric bulge $r_a =r_b$), which is defined as
\begin{align}
\rho_{h}(\mu)=\frac{M}{2 \pi a b c} \frac{1}{\mu(1+\mu)^{3}} && \text{where} && \mu^{2}=\frac{x^{2}}{r_a^{2}}+\frac{y^{2}}{r_b^{2}}+\frac{z^{2}}{r_c^{2}}
\end{align}

{\noindent From} this we get the potential for every position, $x,y,z$, using the method of ellipsoidal shells \citep{chandrasekhar1969,binney2008},

\begin{equation}
\Phi_\mathrm{bulge}(\xi)=-\frac{G M}{2} \int_{0}^{\infty} \frac{d u}{\sqrt{r_a^{2}+u} \sqrt{r_b^{2}+u} \sqrt{r_c^{2}+u}[1+\xi(u)]^{2}}
\end{equation}
where 
\begin{equation}
    \xi^{2}(u)=\frac{x^{2}}{r_a^{2}+u}+\frac{y^{2}}{r_b^{2}+u}+\frac{z^{2}}{r_c^{2}+u}
\end{equation}

The total chain of integration and differentiation processes were computed with the state-of-the-art auto-differentiation capabilities of the Julia language \citep{bezansonJuliaFreshApproach2015} using the ForwardDiff.jl \citep{RevelsLubinPapamarkou2016} and quadgk.jl \citep{quadgk} packages. The equations were solved using the DifferentialEquations.jl package \citep{Rackauckas2017}.

\section{Results}
\label{par:results}

\subsection{ Basic data products}

The integrated \coto\ emission consists mainly of a highly inclined, disk-like structure extending out to 15\arcsec. This distribution is spatially correlated with the dust absorption visible in Figure~\ref{fig:hst}. The main disk structure is clumpy and asymmetric with respect to the center of the galaxy, with 66\% of the total intensity belonging to the southern part of the disk and concentrated in an area $1.3\, \mathrm{kpc}^2$ wide, around $1\kpc$ southwest of the center (30\% of the total intensity). 

The total integrated intensity is estimated to be $S \Delta V = 165 \jykms$, where we did not include flux below three times the local rms noise level. 
We estimate the total line luminosity using the \citet{solomon2005} formula: 
\begin{align}
  L'_\text{CO (2-1)}=3.25\times 10^{7} \frac{ S\Delta V D_L^2}{\nu^2 _\text{CO (2-1)} (1+z)^3}  && [\kkmspc] 
\end{align}
where $\nu$ is the rest-frame \coto\ frequency in GHz, and $D_L$ is the luminosity distance, which for $z=0.014$ corresponds to $D_L = 62.88\Mpc$. This gives $L'_{\rm CO (2-1)}=3.28\times10^{8}\kkmspc$.

From the integrated luminosity, we estimate the total molecular gas mass to be $1.8\times 10^{9} \msun$, using the relation $M_{\htwo} = a_\mathrm{CO} L'_\mathrm{CO}$  \citep{solomon1997}. We adopt a Galactic value of $4.6 \msun (\kkmspc)^{-1}$ \citep{bolatto2013} for the conversion factor, $a_\mathrm{CO}$, and a
\coto\ to \cooz\ flux ratio of four.

Based on the integrated intensity, first moment (mean velocity), and second moment (velocity dispersion) maps of the \coto\ and \cott\ lines (Figures ~\ref{fig:CO21moments} and \ref{fig:CO32moments}), we identify multiple gas structures. In the inner regions, the presence of a bright molecular structure extending up to $650\pc$ is evident, with $i\approx65\deg$ and $\mathrm{PA}=72\deg$, closely aligned with the warm \htwo\ emission \citep{maccagni2016a}. This structure appears disk-like but could also resemble a ring.

\begin{figure}[h]
    \begin{center}
        \resizebox{\hsize}{!}{\includegraphics{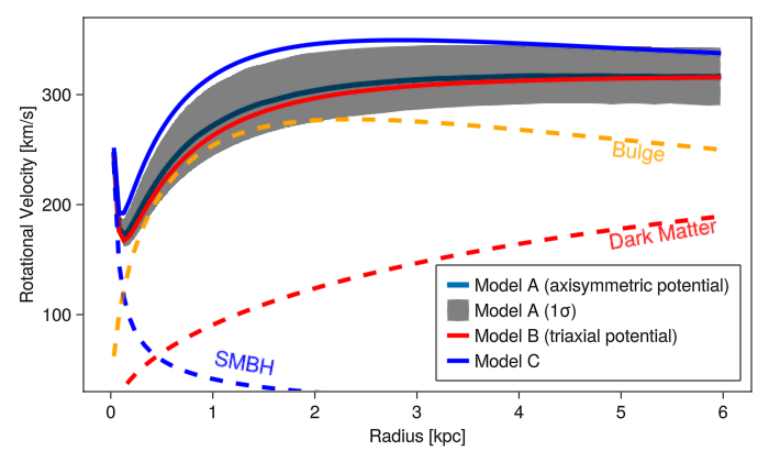}}
        \caption{Rotational velocity. The black line represents the best solution of the axisymmetric model, red the triaxial, and blue the generic interpolating model. The gray band represents the $1\sigma$ confidence level of the parameters' posterior distribution regarding the axisymmetric model.}
        \label{fig:rotation_velocity}
    \end{center}
\end{figure}

The bulk of the molecular gas emission, however, is distributed in a main disk that is oriented north-south and highly inclined. The kinematics observed in the mean velocity map (Figure~\ref{fig:CO21moments}, middle panel) suggests the presence of a warp, as the PA clearly changes from $140\deg$ at $1.4\kpc$ to $180\deg$ at the maximum extent of $4\kpc$. In the southwest portion of the disk, there is a massive substructure spanning an area of approximately $1.3\,\mathrm{kpc}^2$, centered at $\approx 1\kpc$ from the center. This substructure has a mass of around $\sim10^8\msun$ (assuming $a_\mathrm{CO}=4.6 \msun (\kkmspc)^{-1}$) and exhibits the maximum observed velocities (exceeding $350\kms$) and velocity dispersion (with values exceeding $80\kms$). A similar behavior can be observed in the opposite direction (PA $\simeq -30\deg$), though in much smaller and less massive clumps of molecular gas. Both of these regions coincide with notable \htwo\ regions outside of the inner disk \citep{maccagni2016a}.

\subsection{Fitting the ALMA data cube with our 3D code}
\label{sec:warp_model}

We used our new Bayesian code to fit the full ALMA cube and produce the galaxy's rotation curve for several assumptions. In models A and B, we used a physical model that describes the time-dependent evolution of a set of tilted rings within a triaxial axisymmetric modification of the Hernquist potential. The physical approach in models A and B captures the warped disk motion and structure, which results from gas infalling toward the gravitational potential, feeling torques from a nonspherical potential, and dissipative forces. In a third model, named C, we adopted the more generic assumptions mentioned in section~\ref{sec:generic_model}. 
 
For models A and B, there are 12 and 13 fittable parameters, respectively. These include the bulge stellar mass, the scale size on the three axes, $r_a$, $r_b$, $r_c$ (for the axisymmetric model, $r_a=r_b$), the virial mass and concentration parameter of the dark matter component, the initial values of $\theta(0,r)=\theta_0$ and $\lambda(0,r)=\lambda_0$, the settling angle, $\alpha$, the time elapsed from the initial configuration, $\tau$, and the orientation of the reference ring ($i_\mathrm{Bulge}$, $\mathrm{PA}_\mathrm{Bulge}$). The corner plot depicting the posterior distribution of the parameters for Model A is presented in Figure~\ref{fig:corner_ax}.

\begin{table}[!h]
\small
    \caption{Resulting parameters for models A, B, and C.}
    \begin{center}
    \begin{tabular}{ |c| c |c | c|}
     \hline
     Parameter & Model A &  Model B &  Model C \\ 
        \hline
        $M_*$ & $1.7\times 10^{11}\msun$ & $1.6\times 10^{11}\msun$ & $2.1\times 10^{11}\msun$ \\  
        $M_\mathrm{vir}$ & $3.4\times10^{12}\msun$ & $5.4\times10^{12}\msun$ & $2.5\times10^{12}\msun$\\  
        $r_a$ & $2.26\kpc$ & $2.7\kpc$ & $2.1\kpc$\\  
        $r_b$ & $2.26\kpc$ & $2.26\kpc$ & -\\  
        $r_c$ & $2.54\kpc$ & $2.4\kpc$ & -\\  
        $\theta_0$ & $-87.8\deg$ & $-89\deg$ & -\\  
        $\lambda_0$ & $112\deg$ & $114\deg$ & -\\  
        $\alpha$ & $60.6\deg$ & $54.9\deg$ & -\\   
        $c_\mathrm{vir}$ & $7.4$ & $7.4$ & $7.4$\\  
        $\tau$ & $120\Myr$ & $380\Myr$ & -\\  
        $i_\mathrm{Bulge}$ & $61.7\deg$ & $63.8\deg$ & $62\deg$\\  
        $\mathrm{PA}_\mathrm{Bulge}$ & $257\deg$ & $258\deg$  & $257\deg$\\  
         \hline
    \end{tabular}
     
     \tablefoot{$M_*$ is the bulge stellar mass, $r_a$, $r_b$, and $r_c$ are the three axis scales and $i_\mathrm{Bulge}$, $\mathrm{PA}_\mathrm{Bulge}$ is the orientation of the  axis, $c$, relative to the observer. $M_\mathrm{vir}$ is the virial mass and $c_\mathrm{vir}$ the concentration parameter of the dark matter. $\theta_0$ and $\lambda_0$ are the starting values of the disk orientation. $\alpha$ is the dissipation settling angle and $\tau$ the time after the initial values.}
    \label{tab:parameters}
    \end{center}
\end{table}

\begin{figure*}[h]
    \begin{center}
         \includegraphics[width=17cm]{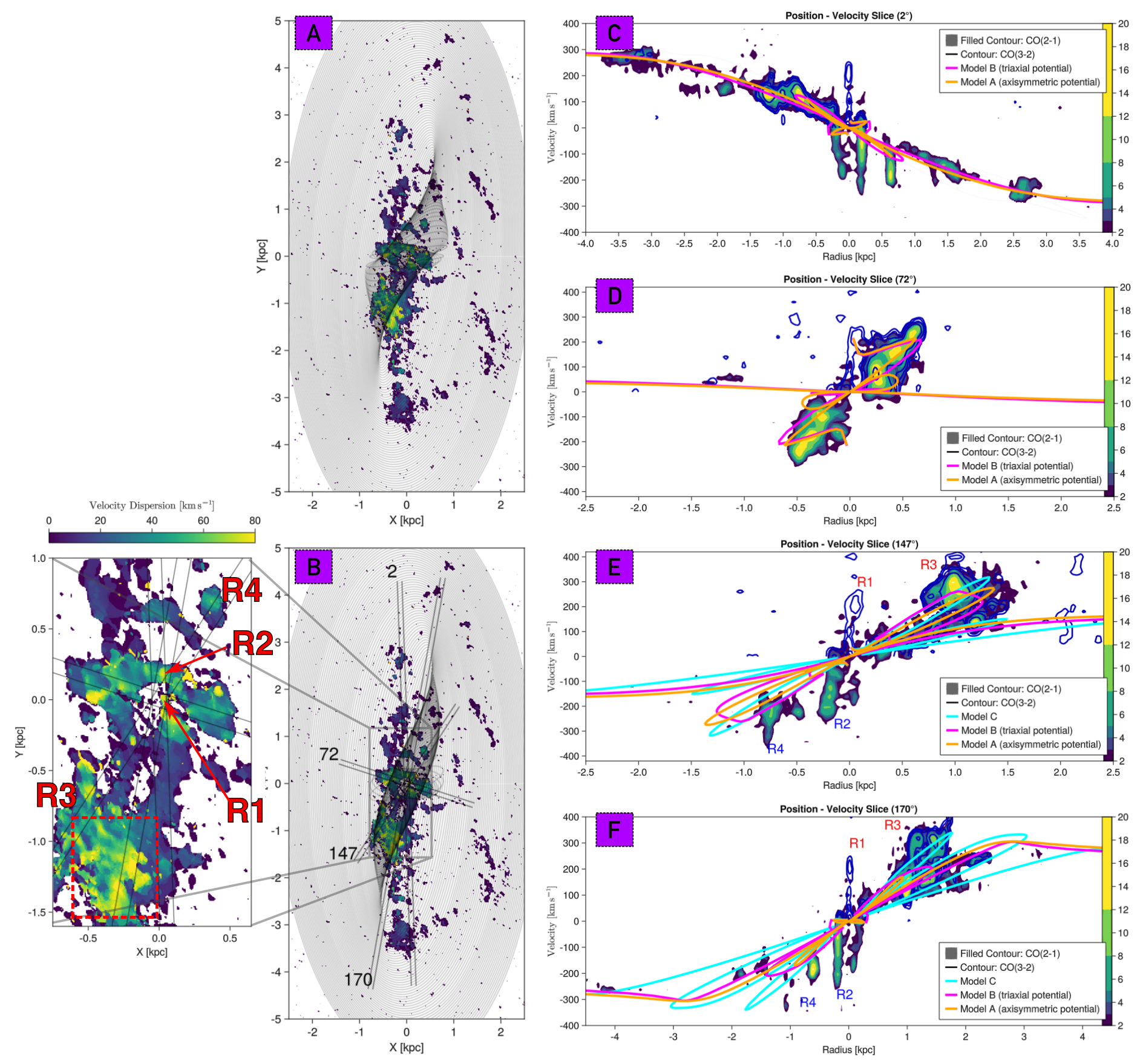}
        \caption{Position velocity slices of CO(2-1) and CO(3-2) emission cubes. Panels \textbf{A} and \textbf{B}: Sky-projected tilted rings for the axisymmetric potential model (A) and the triaxial potential model (B) embedded on top of the CO(2-1) dispersion map. 
        Panels \textbf{C} and \textbf{D}: Slices at angles of $2\deg$ and $72\deg$, covering most of the gas emission and the major axes of the inner disk, respectively. Superimposed on these slices are the velocity projections of models A and B. Model C is omitted to avoid clutter, as its projected velocities are similar to those of the other models. The color map represents the SNR.
        Panels \textbf{E} and \textbf{F}: Position velocity slices along the PAs of $147\deg$ and $170\deg$ where we observe the maximum gas dispersion. Embedded in these slits are also the velocity projections of models A, B, and C.
        }
        \label{fig:pvd}
    \end{center}
\end{figure*}

For model C, $\theta(r)$ and $\lambda(r)$ were generated using an interpolated set of $N=10$ pairs ($r_k$, $\theta_k$, $\lambda_k$), which are fitable parameters. The rotation curve uses the same parameters as the other models, which are the bulge stellar mass, a scale size for the bulge, $r_a$, and the virial mass of the dark matter component, for a total of 26 parameters.

Panels A and B of Figure~\ref{fig:geometry} show the resulting geometry in the galactic plane, characterized by the angles $\theta(t,r)$ and $\lambda(t,r)$, as defined in the previous section.
Panels C and D of the same figure display the inclination and PA of the rings, as observed in the plane of the sky. 

Given the resulting 3D structure, we can observe the molecular gas from an alternative viewpoint (Figure~\ref{fig:3d}), where we see a spiral-like structure as the gas moves toward the central disk. In Figure~\ref{fig:rotation_velocity} we present the rotational velocity and its components. The best-fitting geometrical parameters reveal that the disk has settled into a stable configuration within the central $600\pc$ (the inner disk) $100-400\Myr$ after its initial configuration. Most of the emission observed in the inner disk is a result of the projection of multiple positions of the disk, with the primary contribution coming from the inner disk itself and a smaller contribution from the rapidly warping outer regions. This explains the observed velocity dispersion and the small asymmetry of the emission along the inner disk's major axis. The orientation of the disk rapidly changes over the next $\sim 1\kpc$, with similar changes in the azimuthal angle reproduced by both the axisymmetric and the triaxial models due to the dominant diffusion term. However, the change in the longitude of the ascending node ($\lambda$) shows a different behavior in the two models, with the disk in the triaxial model shifting inward of $1.5\kpc$ to align with a secondary preferred axis. This results in slight differences in the observed inclination and PA curves, which can be seen in the 3D data plane by projecting the results onto several positional velocity slits (as shown in Figure~\ref{fig:pvd}). 

\begin{figure*}[h]
    \begin{center}
        \includegraphics[width=16cm]{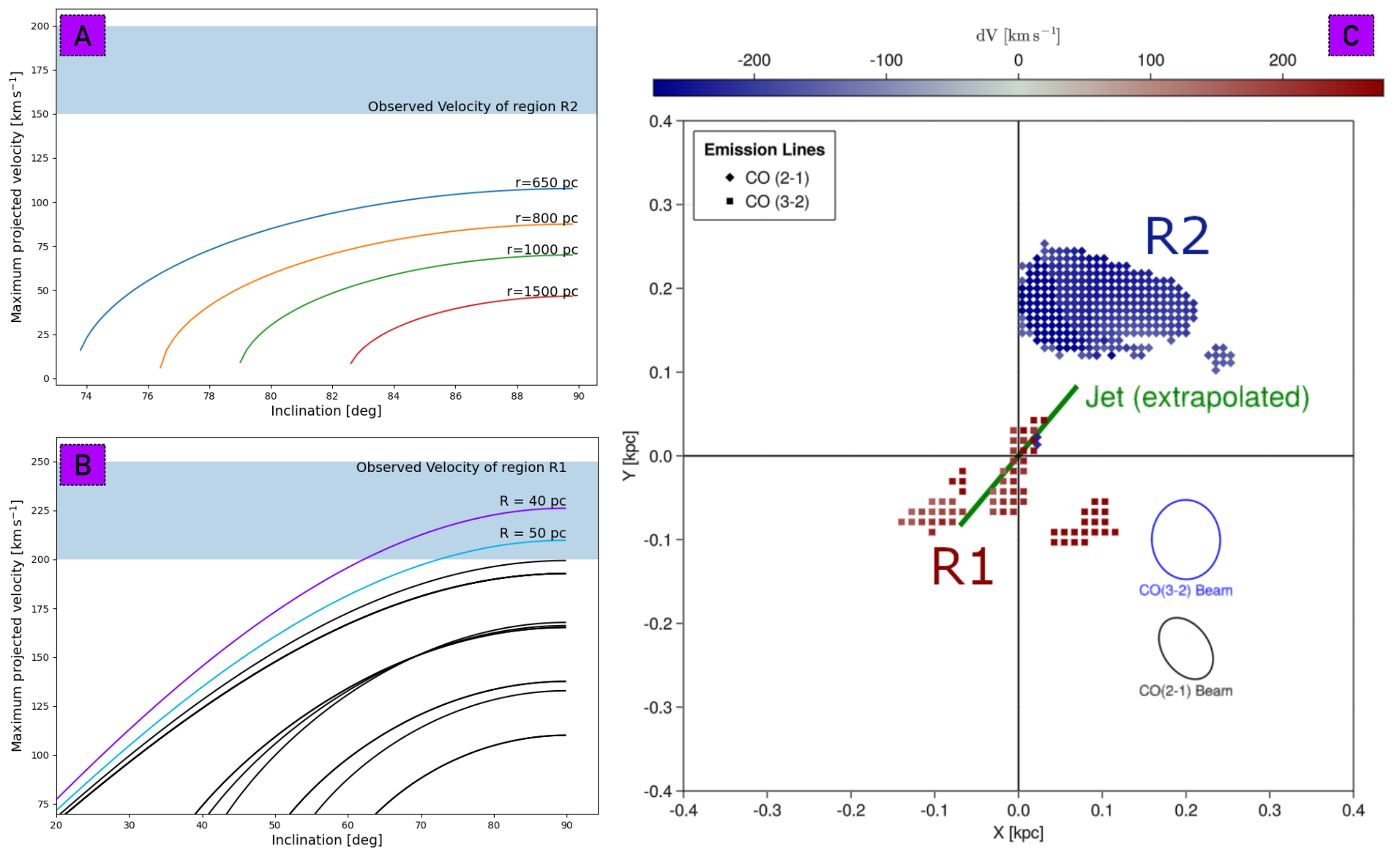}
        \caption{Residuals in the central disk region. \textbf{A}: Maximum projected velocity based on a clump located at a projected radius $R=200\pc$ for different possible orbit radii and inclinations. The blue band represents the observed projected velocity of the \coto\ emission lines.
        \textbf{B}: Maximum projected velocity based on a clump located in the central $R\simeq50\pc$ for different possible orbit radii and inclinations. The blue band represents the observed projected velocity of the \cott\ emission lines.
        \textbf{C}: Emission line residuals ($\delta V>100\kms$) in the central $500\pc$ disk area in the sky plane. Circles represent the \coto\ emission and rectangles \cott\. The green vector is the extrapolated direction of the jet. 
        }
        \label{fig:residuals}
    \end{center}
\end{figure*}

\subsection{Residuals as nonrotating components}
\label{se:residuals}

In all three models, there exist molecular gas regions that cannot be fitted by circular orbits (deviating by a fiducial velocity of $50\kms$ or more from each best-fitting model) and that are thus considered kinematic outliers. Specifically, models A and B fit 80\% of all molecular clouds, while model C fits $94\%$ of them. The residuals correspond to clouds of interest for inflows/outflows, and they are clustered in the four areas (R1, R2, R3, and R4) shown in Figure~\ref{fig:pvd}. 

Two of these regions are located in the central disk region: a set of \cott\ clumps (referred to as \textbf{R1}) at the very center ($\leq120\pc$) and a \coto\ clump located at a projected distance of $R\simeq 200\pc$ in the NW ($\mathrm{PA}=-50\deg$ to $-60\deg$) referred to as \textbf{R2}. Both of these regions exhibit deviations of more than $100\kms$ from the circular velocity model (panel C in Figure~\ref{fig:residuals}). To explore all possible kinematic scenarios for these regions, we first examine whether they could be intrinsically located at larger radii, with projection properties different from those identified by our code -- a scenario that we rule out.

We start with R2, the region that exhibits the highest deviation ($dV\sim -150\kms$), to investigate whether it could be located at kpc scales. Its maximum projected velocity can be estimated using the equation $V_\mathrm{los} = V(r)\frac{\sin i \sin \delta\phi}{\sqrt{1+\tan^2i \cos^2 \delta\phi}}$, where $\delta\phi=\phi -\mathrm{PA}$ is the angular distance between the PA of the orbit and the position of the cloud, and $V(r)$ is the intrinsic rotational velocity. By combining this equation with $R(\phi)$, the sky-projected distance of the orbit with radius $r$, we can compute the maximum projected velocity as

\begin{figure}[h]
    \resizebox{\hsize}{!}{\includegraphics{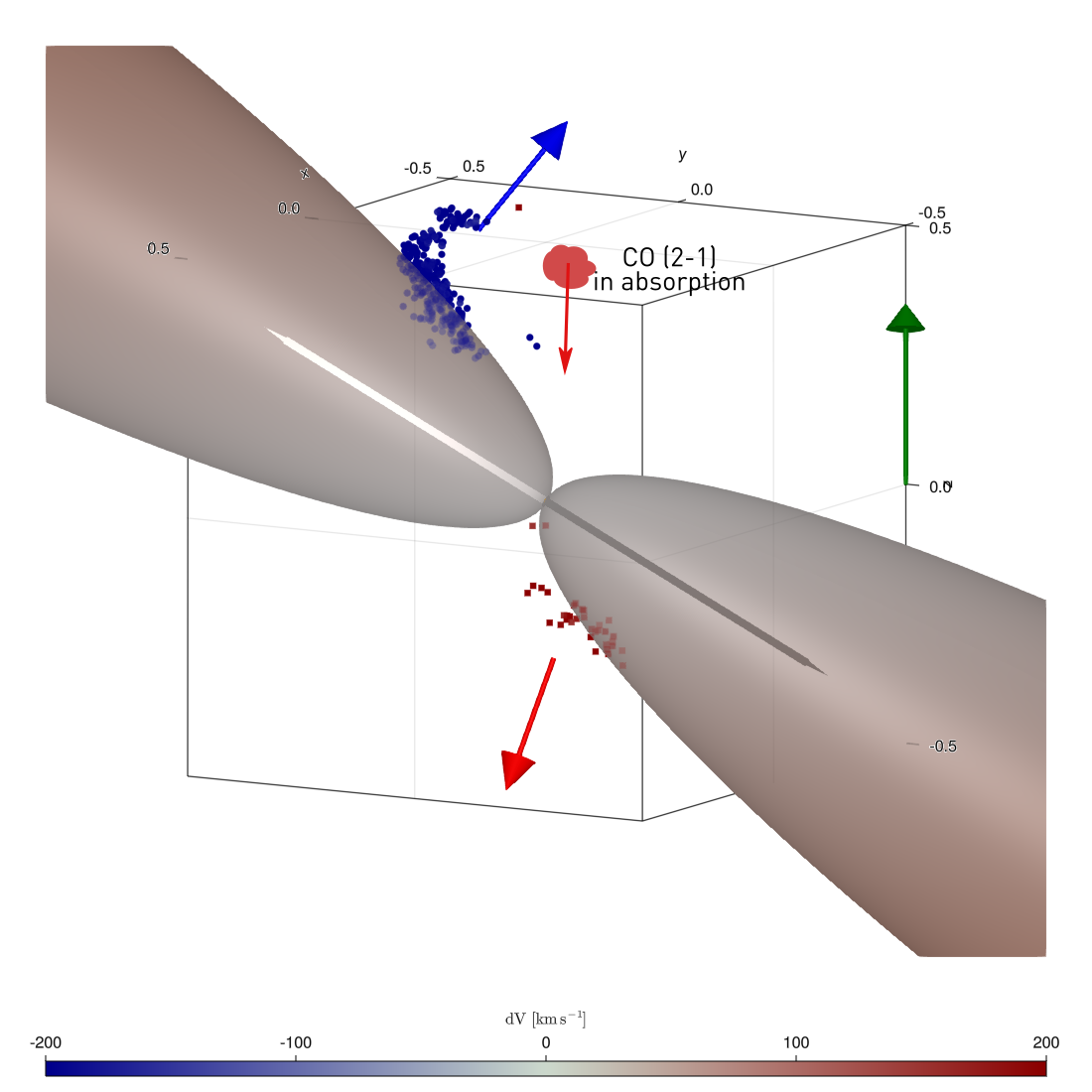}}
    \caption{Emission line residuals ($dV>100\kms$) in the central $500\pc$ disk from another angle. Circles represent the \coto\ emission and rectangles \cott\. The conical shapes represent a hypothetical jet-driven bubble responsible for the molecular gas outflow. The green vector shows the direction of the observer. Blue and red vectors show the possible direction of the observed outflows.}
    \label{fig:inner_outflow_3d}
\end{figure}

\begin{figure*}[h]
    \begin{center}
        \includegraphics[width=18cm]{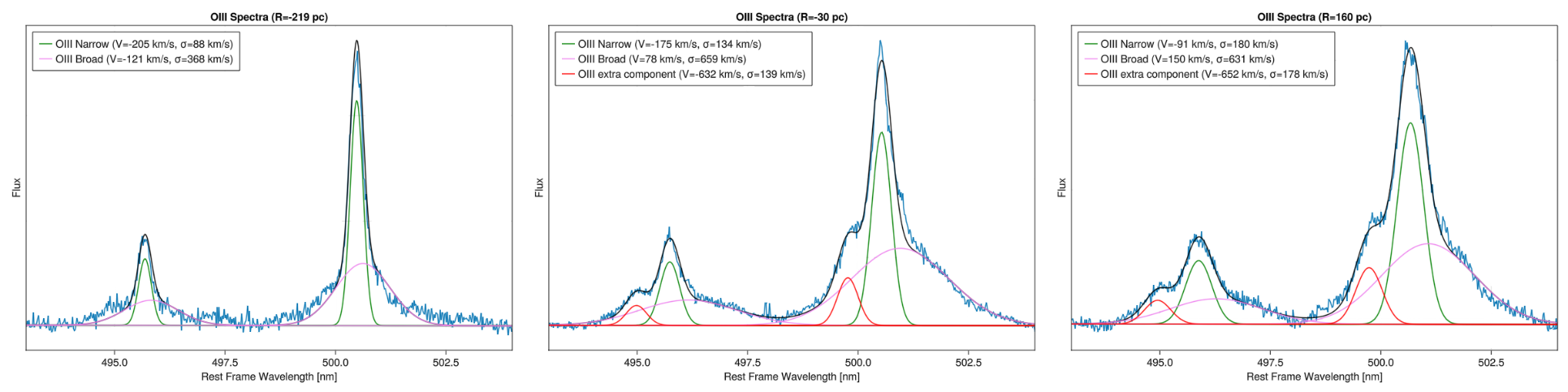}
        \caption{[OIII] spectrum centered in three different positions along a $240\pc$-wide slit (PA $160\deg$). From left to right, $-220\pc$, $-30\pc$, and $160\pc$ from the nucleus. For every position the signal (blue) is averaged out over four pixels, corresponding to a physical width of $142\pc$ along the slit.  In each spectrum, we have fitted one narrow and one broad component. At the $-30\pc$ and $160\pc$ positions, an additional blueshifted ($\approx600\kms$)  component was required to obtain a good fit.}
        \label{fig:OIII_spectra}
    \end{center}
\end{figure*}

\begin{figure}[h]
    \begin{center}
        \resizebox{\hsize}{!}{\includegraphics{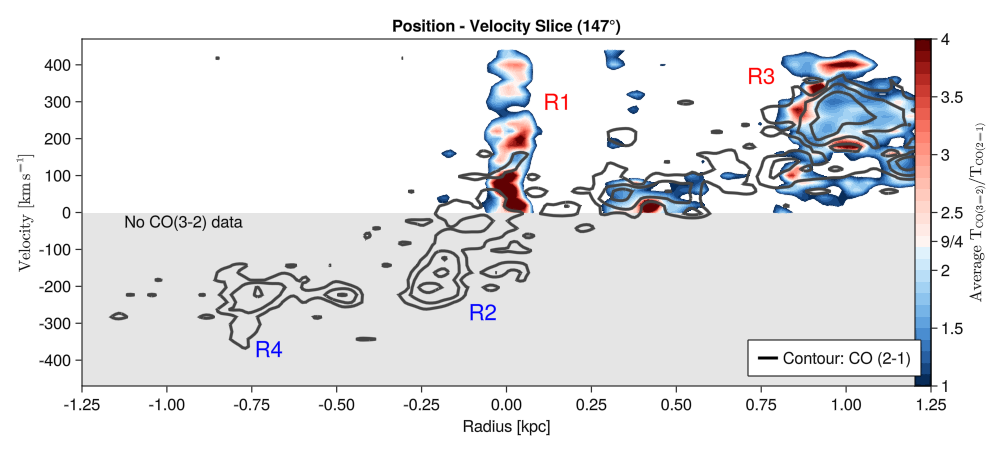}}
        \caption{Position velocity slice along the PA of $147\deg$. The values represent the average value of $T_\coto\ / T_\cott\ $ and the contours the \coto\ intensity.}
        \label{fig:pvd_excitation}
    \end{center}
\end{figure}

\begin{equation}
    \label{eq:maximum_v_los}
    V^\mathrm{max}_\mathrm{los}(R)=V(r)\frac{R}{r}\cos i\sqrt{\tan^2i- \Big(\frac{r}{R}\Big)^2-1} .
\end{equation}

{\noindent For} an intrinsic rotational velocity of $350\kms$ at $r=650\pc$, the maximum projected velocity would be $\sim100\kms$ (panel A in Figure~\ref{fig:residuals}). For larger distances, this value would decrease, indicating that there is no possible circular orbit projection that can explain the observed deviation. Therefore, R2 must be located in the inner disk. Furthermore, its projected velocity ($200-250\kms$) is close to the maximum value that the potential can support (and higher than our best-fitting value). This implies that R2 must be close to the PA of a nearly edge-on orbit, which contradicts the evidence that the rest of the gas at similar radii is moving in the reverse direction and with a different orientation (PA$=72\deg$). Therefore, if R2 is not counter-rotating, it must be in an outflow (or an inflow).

Regarding the high-velocity structure R1, similar conclusions can be drawn. High distances are ruled out by Equation~\ref{eq:maximum_v_los}. Conversely, very small distances ($\leq120\pc$ from the center) and beam smearing (with a  \cott\ beam of $95\pc$) can potentially raise doubts about its central location.
If the gas were regularly rotating and the deviation were induced by a beam smearing effect, then its sky-projected distance from the center must be $R<50\pc$ (panel B of Figure~\ref{fig:residuals}). However, at its maximum extent, R1 reaches a distance of $R\simeq 120\pc$ from the center (panel C of Figure~\ref{fig:residuals}), larger than the major axis of the \cott\ beam ($93\pc$). Thus, we can safely conclude that R1 is also a nonrotating clump. In conclusion, the two central components, R1 and R2, are unlikely to be attributed to a regularly rotating disk, but are instead tracing outflowing (or inflowing) gas. 

Both components exhibit high-velocity dispersion, implying that the gas is either in a highly turbulent dynamical state or that it experiences a large velocity gradient ($>200\kms / 100\pc$) due to acceleration. Previous observations of NGC6328 also indicate the presence of both inflowing and outflowing gas, based on their relative velocity to the source. In front of the radio source, \cite{maccagni2018} observed a CO(2-1) clump in absorption, redshifted at $\sim 340\kms$ and with a velocity dispersion of $\sim 60\kms$. 
Absorption features have also been detected in HI \citep{maccagni2014}. One of them is broad, with a width of $65\kms$, and it is redshifted to a velocity of $26\kms$. The second one is narrower, with a width of $43\kms$, and it is blueshifted to a velocity of $-74\kms$.

Based on the gas kinematic model and the observed dust extinction, we can deduce that the northern side of the inner disk corresponds to its near side (between the center of the galaxy and the observer). Hence, the blueshifted clump is outflowing from the disk. Similarly, the redshifted southern component, R1, on the far side of the inner disk, is also outflowing. This scenario is depicted in Figure~\ref{fig:inner_outflow_3d}.

The outflow scenario is also supported  by two other arguments. The first is the existence of a broad kinematic component at the same locations in the ionized gas emission, as seen in recent optical, X-shooter data. The spectra, extracted from a slit with PA=$160\deg$, are shown in Figure~\ref{fig:OIII_spectra}. They illustrate the 5007 Angstrom [OIII] emission at three different positions along this slit: R2, located $219\pc$ North-East (NE) of the center, the central region at $-30\pc$, and the southwest region at $160\pc$. The spectra at each location are summed over four spectral pixels, and cover a width of $>$100\pc. For all regions, the gas exhibits similar narrow ($\sigma=80-180\kms$) and broad ($\sigma=460-660\kms$) emission line components, and in the last two regions, an extra blueshifted high-velocity component is seen at $\approx 630\kms$. All these findings constitute direct evidence of the presence of strong winds in the inner disk of NGC6328.

The second argument favoring the existence of an outflow is the CO excitation. From the ratio of \cott\ and \coto\ cubes (both re-imaged to a common resolution), we estimated the excitation of the gas. The brightness temperature ratio cube $\mathrm{T}_\mathrm{CO(3-2)}/\mathrm{T}_\mathrm{CO(2-1)}$ significantly exceeds the value for optically thick gas in the local thermodynamic equilibrium (LTE): 

\begin{equation}
    \frac{\mathrm{T}_\mathrm{CO(3-2)}}{\mathrm{T}_\mathrm{CO(2-1)}}=
\frac{I_\mathrm{CO(3-2)}}{I_\mathrm{CO(2-1)}}\Big(\frac{\nu_\mathrm{CO(2-1)}}{\nu_\mathrm{CO(3-2)}}\Big)^2 \overset{\mathrm{LTE}}{=}\frac{9}{4}.
\end{equation}

{\noindent We} observe significantly higher brightness temperature ratios in (and close to) R1, while the bulk of the regularly rotating molecular gas is sub-thermally excited (Figure~\ref{fig:pvd_excitation}). The high excitation ratio could be the result of extra density components, of evaporation of the outer cloud layers that makes them optically thin, or of the presence of extra excitation mechanisms (e.g., shocks, cosmic rays) in the outflow. Similar results for the molecular gas excitation have been obtained in the nearby galaxies IC\,5063 \citep{dasyra2016} and NGC3100 \citep{ruffa2022} and demonstrated to have been induced by an interaction between the radio jets and the surrounding cold gas. These results also suggest that the high-velocity clump in R1 corresponds to a co-spatial and also kinematically deviating warm \htwo\ component \citep{maccagni2016a}.
The other two locations of interest cover part of the high-velocity dispersion regions at kpc distances, to the southwest (\textbf{R3}) and northeast (\textbf{R4}) of the center (as previously mentioned in Section~\ref{sec:ALMAdata}). For these regions, the fraction of fitted clouds shows a significant discrepancy between models A, B ($ 63\%$ fitted), and C ($91\%$ fitted) when introducing a $10\deg-20\deg$ wobble in the $\theta(r)$ curve (panel A of Figure~\ref{fig:geometry}). This wobble can be caused by more than one asymmetry in the total potential, such as stronger bulge triaxiality, nonspherical dark matter potential, a disk, or a bar. Similar wobbles in the polar angle have been observed at $100\pc$ scales in M84 \citep{quillen1999}, where the authors also suggest the presence of torques caused by the AGN radiation pressure. However, the lack of strong observable asymmetry signatures (e.g., bars) in optical or near infrared images raises the possibility that R3 and R4 are also regions of noncircular motions, and -- namely -- outflows.
In R3, some CO-emitting clouds that are more excited than average are also identified. The ionized gas kinematics there do not show a broad emission component, but different velocities than the disk. Leaving the outflow scenario as an open option there, we proceed to the identification of the driving mechanism of the certainly outflowing gas.

\section{Discussion}
\label{sec:discussion}
\label{sec:infall}
Our best models describing the evolution of the majority of the gas disk suggest a recent episode ($100-400\Myr$) of gas accretion. Numerical simulations conducted by \cite{vandevoort2015} predict that a merger can cause such a strong inflow of gas, followed by strong outflows through which the gas can be expelled and re-accreted. These accretion episodes can result in a misalignment between the gas and the galaxy potential, which under the process of gravitational relaxation forms a warped disk due to the gravitational torques of the stellar potential. This process has been recently observed in \object{NGC5077} \citep{raimundo2021} and \object{NGC3100} \citep{ruffa2022,Maccagni2023}, where the misaligned cold gas feeds the SMBH and triggers similar outflows. 

\subsection{What drives the outflow}
\label{se:outflow_energetics}

Energetic and momentum arguments are often used as indicators of the outflow driving mechanism. In order to estimate the mass outflow rate, kinetic power, and momentum, we considered the starting position, velocity, and geometry of the outflow. This is assumed to emanate from a central region and travel outward, so its geometry can be approximated by a cone or sphere. In this context, the term $r_\mathrm{out}$ refers to the distance the outflow has traveled from its origin, representing a measure of how far the material has been carried by the outflow. This simplifying assumption may not always hold, but it is frequently made in the literature.

The outflow mass rate can then be described by the equation
\begin{equation}
    \dot{M}_\mathrm{out}=C  \frac{M_\mathrm{out}V_\mathrm{out}}{r_\mathrm{out}}
    \label{eq:outflow_mass_rate}
\end{equation}
where $M_\mathrm{out}$ and $V_\mathrm{out}$ represent the mass and velocity of the outflow, respectively, and $C$ is a parameter that depends on the outflow history \citep{lutz2020}. We use a value of $C=1$, which assumes that the outflow we are observing now has started at a point in the past and continued with a constant mass outflow rate until now. Another value used in the literature is $C=3$, which takes into account that the outflowing gas has a constant average volume density in a cone or a sphere. For the outflow travel distance, we assumed an indicative value of $r_\mathrm{out}\simeq100\pc$, as this is a typical distance if the outflow has originated locally from a nearby part of the disk (instead of in the center of the galaxy). For the outflowing mass, we used a conversion factor of $a_\mathrm{CO}=0.8 \msun (\kkmspc)^{-1}$. This value is lower than that typically adopted, as we made the assumption that both R1 and R2 have optically thin parts due to interaction with the AGN or jet \citep{cicone2014a,dasyra2016,ruffa2022}.

\begin{figure*}[h]
    \begin{center}
    \includegraphics[width=18cm]{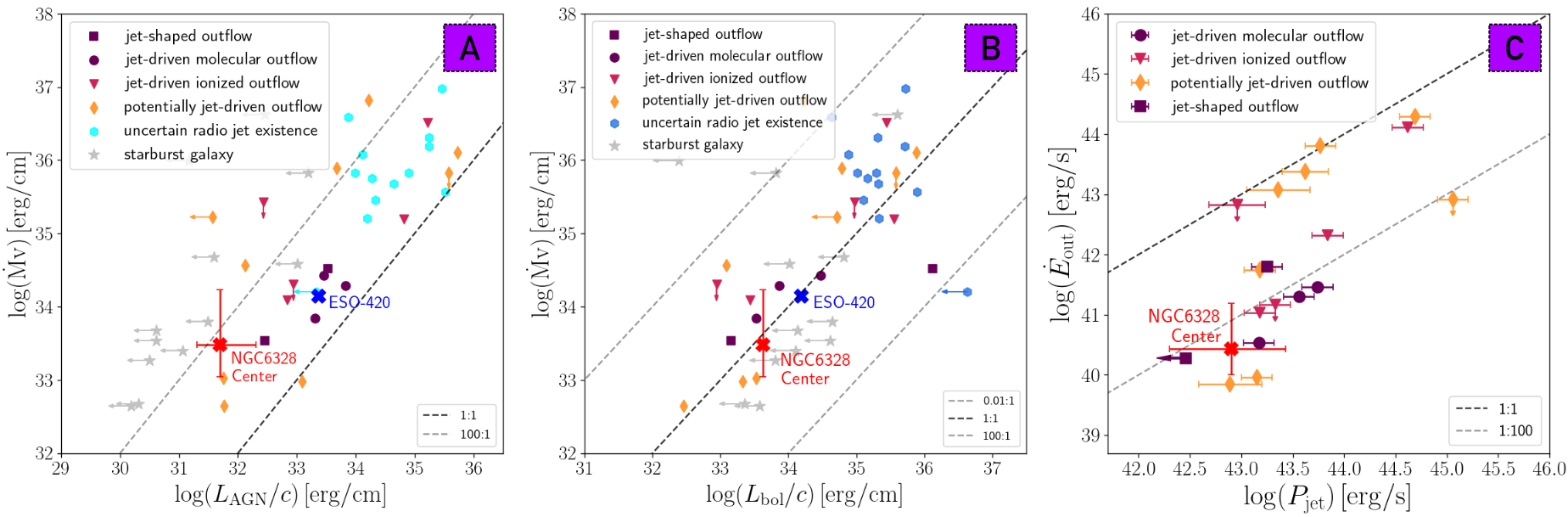}
    \caption{Wind vs. driving mechanism energetics for local galaxies with massive molecular wind detections. \textbf{A}: Wind momentum rate vs. radiation pressure of AGN photons.
    \textbf{B}:Wind kinetic luminosity vs. jet power. The dashed lines indicate different ratios of the plotted parameters. 
    \textbf{C}: Outflow's momentum rate relation with the total photon momentum output of the galaxy. The dashed lines denote the different values of the ratio of outflow's momentum rate over bolometric luminosity. The arrows denote upper limit values.
    }
    \label{fig:literature_AGN_Jet}
    \end{center}
\end{figure*}

To calculate the outflow velocity, $V_\mathrm{out}$, we used the velocity deviating from the best-fitting model. However, this value needs to be corrected for the outflow angle. We tried two different approaches in estimating $V_\mathrm{out}$, first by using the estimation of $V_\mathrm{out}=|\Delta v|+0.12\sigma_V$ proposed by \cite{lutz2020}, where $\sigma_V$ is the velocity dispersion of each cloudlet (emission line per pixel) in the outflow. Then, we used a projection angle based on a uniform distribution from $-30$ to $30\deg$ from the disk and adopted the mean value. This represents all the possible angles in a cone centered to the disk itself with a maximum opening toward the observer. Both methods provided similar results, which are presented in Table~\ref{tab:outflows} and compared to molecular outflows detected in the literature in Figure~\ref{fig:literature_AGN_Jet}. In the event that the outflow originated from the center of the galaxy and had a constant average volume density (i.e.,\,$C=3$), these results would only be a factor of two higher.

\begin{table}[!ht]
\small
\caption{Molecular outflow properties}
\begin{tabular}{ |c| c |c |c |c |c|}
     \hline
     Region & Mass &  $\dot{M}_\mathrm{out}$  & Kinetic Power & Momentum Rate\\ 
      \hline
       & $10^6\msun$ &  $\mathrm{M}_\odot \mathrm{yr^{-1}}$  & $10^{40}\ergs$ & $10^{33}\,\mathrm{dyn}$\\ 
        \hline
     R1 & $0.12$ & $0.2$ & $0.3$ & $0.23$ \\  
     R2 & $1.9$  & $2.9$ & $2.4$ & $2.8$ \\
      \hline
    Total & $2.02$ & $3.1$ & $2.7$ & $3$ \\
     \hline
\end{tabular}
\label{tab:outflows}
\end{table}

We are able to estimate the likelihood of the observed outflow being driven by energy injection from local SF --supernovae (SNae) -- or from the AGN. Based on the formula presented in \cite{sullivan2006}, the rate of Type Ia SNae is given by $\Gamma = 5.3\times 10^{-3} \frac{M}{10^{11} \msun} + 3.9 \times 10^{-4} \frac{\mathrm{SFR}}{\mathrm{M}\odot\,\mathrm{yr}^{-1}}\,\mathrm{SN}\,\mathrm{yr^{-1}}$. This leads to an estimated power output of

\begin{equation}
    \frac{\dot{E}_\mathrm{SN}}{\mathrm{erg s^{-1}}}=1.68\times 10^{41} \frac{M}{10^{11} \msun} + 1.24\times 10^{40} \frac{\mathrm{SFR}}{\mathrm{M}_\odot\,\mathrm{yr}^{-1}}
\end{equation}

Our estimate of the total stellar mass within a radius of $200\pc$, as derived from our modeling, is $10^9\msun$ (or 0.6\% of the total stellar mass). Using this value and assuming that the SF rate (estimated to be $1-2\,\mathrm{M}_\odot \mathrm{yr^{-1}}$, according to \citealt{willett2010}) is proportional to the stellar mass distribution, we estimate the SNae's kinetic power to be $1.8\times10^{39}\ergs$. This is an order of magnitude below the lower limit of the outflow total kinetic power. This suggests that the power from SNae alone is not the main driver of the outflow.

On the other hand, for the low-luminosity AGN ($L_\mathrm{AGN}/c\approx 4.3\times10^{31}\,\mathrm{dyn}$), the kinetic energy ratio of $\dot{E}_\mathrm{out}/L_\mathrm{AGN}\sim0.02$ is consistent with the scenario of a wind driven by the AGN radiation pressure. However, the momentum rate ratio of $\dot{p}_\mathrm{out}/L_\mathrm{AGN}/c\sim 60$ is large even for an intermittent AGN episode scenario, as noted in \cite{zubovas2020}.

The power of the radio jet in NGC6328 has been estimated to be $2\times 10^{42}\ergs$ through SED fitting \citep{sobolewska2022}. However, the use of the calorimetric scaling relation involving the $L_\mathrm{1.4GHz}$ luminosity from \cite{willott_emission_1999}, \cite{birzan2008}, and \cite{Cavagnolo2010} yields a similar value of $2-3\times 10^{43}\ergs$, which is an order of magnitude higher than the estimate obtained from SED fitting. According to \cite{wojtowicz2020}, the minimum jet power is estimated to be $2.7\times 10^{42}\ergs$ under the assumption of energy equipartition between the magnetic field and radio-emitting electrons within the lobes. Even with this minimum estimate, it is still clear that the jet has more than enough energy to drive the outflow and play a significant role in boosting or supporting the AGN's overall impact on its surroundings. Numerical hydrodynamic simulations by \cite{mukherjee2016} have shown that low-powered jets can have a greater impact on the ISM due to their prolonged interaction and difficulty in escaping into intergalactic space.

\subsection{Why can we not see the jet?}
\label{sec:hidden_jet} 
 The fact that the outflow observed in NGC6328 is further away from the nucleus than the currently detected radio emission is consistent with other sources where molecular winds trace previously undetected radio jets, as seen in cases such as \object{NGC1377} \citep{aalto2016}, \object{4C31.04} \citep{zovaro2019}, and \object{ESO420-G13} \citep{fernandez-ontiveros2020}. 

One potential explanation for the lack of radio emission in NGC6328 could be due to recurrent jet-launching powered by stochastic feeding. In this scenario, the energy from previous jets may have been deposited onto large bubbles of kpc scales a long time ago, leading to the cooling of cosmic rays in these bubbles, and thus to the lack of detected radio emission. Using the same assumptions of the outflow properties, it is estimated that the outflow originated $t_\mathrm{out}= 200-700\kyr$ ago (depending on the possible values of $r_\mathrm{out}$). The magnetic field at pc scales is estimated to be $B\approx 10^{-2}\,\mathrm{G}$ \citep{tingay1997}, and the synchrotron cooling time at $0.1\kpc$ scales is computed to be $t_\mathrm{cool,syn} \approx 150\kyr$. This cooling time could be significantly reduced if inverse Compton (IC) cooling were taken into account or if the magnetic field during the previous cycle of the jet were significantly higher. By comparing these two timescales, it can be argued that a highly magnetized relic of a turned-off jet can fade away within an AGN duty cycle due to synchrotron cooling, shock-induced cooling, and adiabatic losses. Such a jet would have been capable of providing mechanical and radiative feedback to the intergalactic medium up to kpc scales. In the case of a mass-loaded jet, in which hadronic radiative processes are important \citep{reynoso2011,rieger2017}, cavities will be observed in cold molecular gas due to photo-dissociation by X-ray and gamma-ray photon fields emanating from the jets \citep{maloney1996,Wolfire2022}.

Another scenario that could explain the lack of radio emission in NGC6328 is the relativistic de-boosting. This occurs when the radio emission from the jet is directed away from the observer, reducing its apparent luminosity and making it more challenging to detect. To assess the impact of relativistic de-boosting, we used radio imaging and the velocity of the radio blobs, as determined through VLBI (estimated to $\beta_\mathrm{app}=0.13\pm0.06$ from \citealt{angioni2019}), to calculate the jet/counter-jet flux density ratio, which we estimated to be $R=1.4\pm0.1$ at 8.4 GHz. The intrinsic $\beta$ and jet inclination, $\theta_\mathrm{jet}$ can estimated by solving the set of equations,
\begin{align}
    \beta_{\mathrm{app}}=\frac{\beta \sin \theta}{1-\beta \cos \theta} &&  R=\left(\frac{1+\beta \cos \theta}{1-\beta \cos \theta}\right)^{p-\alpha}
\end{align}
where $\alpha=-1.5$ is the spectral index and $p$ is a parameter that takes the values two or three depending on whether $R$ is calculated from an integrated flux across a continuous jet or a single component \citep{angioni2019}. Assuming $p=3$ for a single moving component that dominates the 8.4 GHz core radio flux, we estimated $\beta=0.14\pm0.05$ and $\theta_\mathrm{jet}=68\deg\pm17\deg$. 
These results are robust with regard to variations in the $p$ parameter; using $p=2$ yields similar values within the error margins, with $\theta$ having a central value of $62\deg$. Based on these parameters, the Lorentz factor, $\Gamma$, is approximately $1.02$, and the corresponding relativistic boost factor $\delta\sim1.02$. Given the modest value of $\delta$, it is unlikely that relativistic de-boosting is responsible for obscuring the jet from detection.

However, jet acceleration and collimation in AGN have been observed to occur over distances of $10^4-10^6 R_s$ \citep{Blandford2019,kovalev2020,Okino2022,baczko2022,Roychowdhury2023}, which corresponds to more than $0.5\kpc$ for a black hole with a mass of $4\times10^8\msun$. This length scale is similar to the size of the radio structure ($\sim 1\kpc$) observed by VLBI in the present case. Assuming an inclination angle of $\approx 65\deg$ suggests that under sufficient acceleration ($\Gamma > 5$ a typical value for relativistic jets) the jet could reach a de-boosting factor of $\delta^4 \gtrsim 0.01$, making it too faint to be detected by radio telescopes. Additionally, the decrease in magnetic field strength (roughly $r^{-1}$ for a collimated jet; \citealt{vlahakis2004,fendt2006,porth2011}) could further reduce the radio luminosity and explain the lack of radio flux beyond $100\kpc$. At the same time, efficient IC cooling due to an abundance of ambient photons could explain the extended X-ray \citep{beuchert2018} and gamma-ray \citep{migliori2016} components observed in the source spectrum, and the suppression of any extended radio emission by insufficient synchrotron cooling. The presence of a hidden, extended jet would offer greater kinetic power than the current estimations through SED fitting ($\approx 10^{43}\ergs$), and could transfer energy over kpc scales, possibly contributing to the observed excited warm \htwo\ \citep{maccagni2016a} and extended soft X-ray component \citep{beuchert2018}. 

\subsection{The jet-outflow connection in nearby galaxies with massive outflows of cold molecular gas}
\label{sec:literature_jets}

Motivated by the possibility of a jet-ISM interaction in NGC6328, we evaluate whether the role of jets as drivers of gas outflows has been underestimated due to the criterion typically employed for this evaluation. The evaluation primarily involves a parameter, $q_\mathrm{ir}$ \citep{ivison2010}, defined as the logarithmic ratio of the observed far-infrared flux to the $1.4\GHz$ radio flux. In starburst galaxies, these two quantities correlate due to the synchrotron and dust emission around SNae \citep{helou1985}. Consequently, $q_\mathrm{ir}$ falling below a critical value (indicating radio excess) has been interpreted as a criterion of radio loudness \citep{ivison2010}. In a recent compilation of 45 local galaxies with massive molecular outflows, most galaxies were not radio loud based on this criterion and, consequently, jets were deemed to be a rare outflow-launching mechanism \citep{fluetsch2019}. 

However, the discovery of galaxies with jet-shaped outflows but without (strong) associated radio emission \citep{aalto2020,fernandez-ontiveros2020} demonstrates that the radio excess is an unsafe criterion. Outflows can remain once the jets that initiated them fade or move away. We thus reevaluated the jet-outflow connection in these sources, primarily by spatial distribution criteria and secondarily by energetic criteria (see Appendix~\ref{sec:literature_jets}). We find that a plausible jet-outflow connection exists in about half of the examined AGN, as the jet coincides with or extends in the same direction as the outflow. This is true for almost all AGN of low luminosity ($\lesssim10^{11}\,\mathrm{L_\odot}$). From an energetic viewpoint, the jet can sustain the outflow in nearly all spatially identified cases, as the jet power (cavity and synchrotron included) exceeds the outflow kinetic energy. The impact of jets on the molecular ISM could therefore concern a high fraction of local AGN with massive outflows.  

\section{Conclusions}

We present a novel tilted-ring code capable of fitting ALMA cubes in a Bayesian framework, by inferring the intrinsic 3D distribution of the gas. The code incorporates a differential precession model in a triaxial axisymmetric potential (e.g., that of a nonspherical Hernquist bulge), and additionally offers a model that treats rings as independent entities, with interpolation between them to derive their geometrical properties. When applied to \coto\ and \cott\ ALMA data of the central $4\kpc$ of NGC6328, it indicated that the cold gas has settled into a stable inner disk  in the central $600\pc$ $100 - 400\Myr$ on from its initial configuration. The orientation of the disk rapidly changes over the next $1-2\kpc$, explaining the observable high dispersion regions in these areas. We also detect the presence of outflowing molecular gas in at least two (and up to four) regions in the central $200\pc$ of the galaxy. From an energetic viewpoint, the jet is the most plausible mechanism that can sustain the outflow, adding NGC6328 to the catalog of jet-ISM interacting sources where the jet is not detected in the outflow region. We have discussed two possible mechanisms of how this can happen, due to either recurrent jet-launching powered by stochastic feeding or relativistic de-boosting of the jet brightness in the event that the jet becomes collimated after the $2\pc$ that is currently visible.

\begin{acknowledgements}
This project has received funding from the Hellenic Foundation for Research and Innovation (HFRI) and the General Secretariat for Research and Technology (GSRT), under grant agreement No 1882. JAFO acknowledges financial support by the Spanish Ministry of Science and Innovation (MCIN/AEI/10.13039/501100011033) and ``ERDF A way of making Europe'' though the grant PID2021-124918NB-C44; MCIN and the European Union -- NextGenerationEU through the Recovery and Resilience Facility project ICTS-MRR-2021-03-CEFCA. IR acknowledges support from grant ST/S00033X/1 through the UK Science and Technology Facilities Council (STFC).

The figures of this paper have been created thought Julia Packages of Makie.jl \citep{DanischKrumbiegel2021} and Python's Matplotlib \citep{Hunter:2007}. Corner Plots are created with PairPlots.jl \citep{pairplots}
This paper makes use of the following ALMA data: ADS/JAO.ALMA\#2015.1.01359.S, and ADS/JAO.ALMA\#2017.1.01638.S. ALMA is a partnership of ESO (representing its member states), NSF (USA) and NINS (Japan), together with NRC(Canada) and NSC and ASIAA (Taiwan), in cooperation with the Republic of Chile. The Joint ALMA Observatory is operated by ESO, AUI/NRAO and NAOJ. The National Radio Astronomy Observatory is a facility of the National Science Foundation operated under cooperative agreement by Associated Universities, Inc.

\end{acknowledgements}

\bibpunct{(}{)}{;}{a}{}{,} 
\bibliographystyle{aa} 
\bibliography{References.bib} 

\begin{appendix}

\section{Clump detection}
\label{sec:clump_detection}
We implemented a source detection process that encodes the data cube into a set of $M$, individual clumps with sky projected coordinates $\{(x_m,y_m,V_m)\}_{m=1}^M$, through a source detection process. The first step in the process was iterating over all the spaxels of the cube and fitting $N$ Gaussian-shaped emission lines per spaxel $(x,y)$ using the equation
\begin{equation*}
    S^{x,y}(V,\{V_c\}_{i=1}^N,\{F\}_{i=1}^N,\{\sigma\}_{i=1}^N)
=\sum_{i=1}^{N} F_i\exp \Big(-\frac{(V-Vc_i)^2}{2\sigma_i^2}\Big)
\end{equation*}

The number of Gaussian components, $N$, was determined by using the Bayesian information criterion (BIC) as a model comparison criterion that balances the goodness of fit and the number of model parameters. Assuming that the likelihood of observing the signal on each spaxel is a Gaussian with a root mean square noise, $\sigma_F$, the BIC was defined as 
\begin{equation*}
    \mathrm{BIC}=\chi^{2}+ k\log(n)+n\log(2\pi\,\sigma_F^2)
\end{equation*}
where $k$ is the number of fitted parameters, which in our case is $k=3\times N$, and $n$ the size of the signal in one spaxel. Adding more Gaussian components to the emission line model can lead to overfitting of the data, whereby the model becomes too complex and captures the noise. To prevent overfitting, we stopped adding more components to the model when the BIC was minimized.

After the process of fitting multiple lines-sources per spaxel was finished, we merged them together into single clumps by starting from the brightest source and following the dropping flux to their neighbors. This methodology is similar to the astronomical dendrograms algorithm that we also tested by using the astrodendro package in Python \citep{astrodendro}.

The reason we introduced our new methodology rather than directly using the similar estimated clumps from astrodendro is that we wanted to use physically motivated constraints to the fitting procedure. For example, the dendrogram algorithm can combine fluxes to the same clump ("leaf") with a final velocity dispersion of $\sim 80\kms$ or with sizes of a few kpc as the data is smooth enough to support it. As our intention was to fit a warped disk, assuming typical molecular clumps properties, we added constraints supporting our scenario. However, the comparison of the two methods gives similar results.

\section{Transformation of sky coordinates}
\label{sec:Sky_coordinates_transformation}

In order to transform the positions of gas clouds from galactic to observed sky coordinates, a transformation matrix was needed. This matrix was constructed by using a known vector in the galactic coordinates, $\hat{N}_g$, and its counterpart, which serves as a reference ring in the sky coordinates, $\hat{N}_s$:

\begin{equation}
\mathbf{R}=\mathbf{I}+\mathbf{U}+\frac{\mathbf{U}\cdot\mathbf{U}}{1+\hat{N}_g\cdot\hat{N}_s}
\end{equation}

where $U$ is the skew-symmetric cross-product matrix of $u=\hat{N}_g\times\hat{N}_s$,
\begin{equation}
U=\left[\begin{array}{rrr}
0 & -u_{3} & u_{2} \\
u_{3} & 0 & -u_{1} \\
-u_{2} & u_{1} & 0
\end{array}\right]
\end{equation}

As the disk settles toward the preferred axes of the stellar potential, we selected the reference ring to mirror these axes, thereby setting $\hat{N}g=\hat{z}$ (in the galaxy coordinates). The same vector in the sky coordinates corresponds to the vector defined by two angles, $i_\mathrm{Bulge}$ and $\mathrm{PA}_\mathrm{Bulge}$, as given by

\begin{equation}
\hat{N}_s=\begin{pmatrix}
    \sin i_\mathrm{Bulge} \cos\mathrm{PA}_\mathrm{Bulge} \\
    \sin i_\mathrm{Bulge} \sin\mathrm{PA}_\mathrm{Bulge} \\
    \cos i_\mathrm{Bulge}
\end{pmatrix}
\end{equation}

\section{Literature review}

To evaluate the fraction of galaxies that our results concern, we gathered literature-based evidence for links between jets and local galactic outflows. For this purpose, we used a recent compilation of local galaxies with molecular outflows \citep{fluetsch2019}, which we expanded with four more galaxies that had a molecular outflow detection (NGC6328, NGC1377, NGC3100, and ESO420-613). Instead of the radio excess as a criterion for the examination of the jet-outflow connection, we used the following classification scheme for the AGN in the sample: 1) sources with a jet that accelerates molecular gas along its trajectory, 2) sources with a jet that accelerates atomic gas along its trajectory and that can then entrain molecular gas in a multiphase wind model \citep{zubovas2012}, 3) sources with a jet that could be related to the gas acceleration, as it spatially coincides with or extends in a similar direction to the outflow.

 The first group of sources comprises \object{IC5063}, \object{M51}, and \object{NGC1068}. In these galaxies, the jet is associated with part of the molecular gas acceleration, with accelerated blobs and outflow starting points along the jet path. IC5063, which has at least four outflow starting points besides the nucleus, is the prototype of this category \citep{dasyra2015}. In the Seyfert 2 galaxy M51, outflowing molecular blobs are detected near the radio jet outline and the gas excitation changes as a function of distance from the jet \citep{querejeta2016, matsushita2015}. In NGC1068, molecular gas is being entrained at known regions of jet-cloud interaction along the spatially resolved radio jet. This jet-launched outflow adds to the mass of a wind launched by the AGN accretion disk \citep{garcia-burillo2019}. NGC1377 and ESO420-613 also belong to this category, with a jet-shaped outflow that probes the synchrotron emission. The radio-quiet galaxy \object{NGC1377} hosts a molecular outflow that is spatially resolved and jet-shaped, with a width of only $\sim5-20\pc$ along a length of $150\pc$. The outflow revealed the existence of a radio jet that remained elusive in radio images \citep{aalto2020}. In the Seyfert 2 galaxy ESO420-613, ALMA CO data of $30\pc$ resolution reveals the existence of a massive outflow far from the nucleus and along the minor axis with a jet-trailing shape: filamentary emission precedes bifurcated emission at a presumed jet-ISM impact point \citep{fernandez-ontiveros2020}, as in IC5063. Weak radio emission encompasses the outflow location, but its intrinsic shape is unknown as the resolution of the data is considerably lower than that of ALMA.
 
In the second group, the jet has been proven to accelerate atomic gas, with wind starting points along its path. A molecular outflow also exists in these sources, but its connection with the jet remains to be proven. Still, in the current view of multiphase outflows, the passage of an ionized wind by molecular clouds initiates a molecular outflow, as ram pressure and hydrodynamic instabilities peel off the outer cloud layers. Any wind is thus likely to have both atomic and molecular content. A prototype for this category is \object{Mrk1014}, in which an ionized oxygen wind, seen in \oiii, exists at a radio hot spot $\sim3\kpc$ away from the nucleus \citep{fu2009}. An OH outflow, unresolved at these scales, also exists at the nucleus \citep{veilleux2013}. In 4c12.50, atomic hydrogen is seen in absorption in front of the radio-jet tip \citep{morganti2004}. A nuclear, high-velocity CO outflow, again unresolved at the radio-jet scales, is also detected \citep{fotopoulou2019}. In \object{NGC1266}, high-velocity H$\alpha$ is seen inside and near the spatially resolved jet, probed by its 1.4 and 5GHz emission \citep{davis2012}. A nuclear molecular outflow, offset in positive and negative velocities along a similar orientation was detected in CO \cite{alatalo2011a}. In \object{NGC5643}, an ionized outflow delineates the jet axis \citep{cresci2015}, while a potential CO detection \citep{alonso-herrero2018} is here treated as an upper limit.

The third group comprises sources in which the jet could be related to the outflow, because it spatially coincides with or extends in the same direction as the outflow. In \object{NGC613}, which is a spiral galaxy with a resolved radio jet and a nuclear outflow with an extremely high momentum rate ($\sim 400 L_\mathrm{AGN} /c$), an offset between the blueshifted and redshifted CO emission is seen in the jet direction \citep{audibert2019}. 
In the merging system \object{NGC3256}, the southern nucleus shows a disk plus a bipolar jet radio map; an outflow is located at both jet edges \citep{sakamoto2014a}. Other galaxies with molecular outflows in the direction of large or small-scale jets comprise \object{NGC1433} \citep{combes2013}, \object{NGC6764} \citep{leon2007}, Circinus \citep{zschaechner2016,harnett1987}, and \object{Mrk273} \citep{cicone2014a,gonzalez-alfonso2017}. 
In \object{PKS1549-79}, an \oiii\ wind fills a trajectory gap between a massive CO nuclear outflow and the radio jet \cite{oosterloo2019}. In \object{HE1353-1917}, a spectacular \oiii\ and $H_{\alpha}$ wind that leaves the galactic disk at the point of a jet-cloud interaction could have a molecular counterpart, which we treat as an upper limit \citep{husemann2019}. In \object{3C273} the \oiii\ kinematics show counter-motions with respect to the disk along the jet trail; a molecular arc perpendicular to the jet trail is associated with the cocoon \cite{husemann2019a}. We again treat it as an upper limit. Another source that can be included in this third group is NGC3100 \citep{ruffa2022}, which is a lenticular galaxy with young, resolved radio jets (total linear extent $\approx2\kpc$). Here outflowing, optically thin molecular gas with very high excitation temperatures (i.e.,\,$T_{\rm ex}\gg50$~K) is detected around the jet path, indicating that an ongoing jet-cold gas coupling is altering the physics and kinematics of the involved gas fraction.

In Figure~\ref{fig:literature_AGN_Jet}, we plot the outflow kinetic energy vs. the jet power, as well as the outflow kinetic luminosity vs. the AGN radiation pressure, color coding galaxies based on the above-mentioned classification scheme. Outflow and galaxy properties are summarized in Table~\ref{table:literature}. Outflow masses are consistently computed using an \xco\ of $4\times10^{19}\,\mathrm{cm^{-2}(K km s^{-1}})^{-1}$ in all galaxies but IC5063. We find that in nearly all low-luminosity AGN there is either a certain or a plausible link between the jet and the outflow. The only exception is NGC4418, for which the presence of a radio jet is unclear (but cannot be ruled out) based on its clumpy and elongated 5GHz morphology \citep{varenius2014}. Even at high AGN luminosities, certain galaxies have unambiguously jet-driven outflows. In all (low- and high-luminosity) AGN, momentum boosting would have been required to explain the observed outflow if it were originating from the AGN radiation pressure alone \citep{cicone2014a}. The energy deposition by the jet softens or revokes the need for momentum boosting, which occurs during an adiabatic expansion of a radiatively driven bubble of an AGN. It also reduces the need for alternative explanations to momentum boosting such as AGN variability. In the variability scenario, the ongoing outflows were driven by radiation during earlier, more luminous phases of AGN activity \citep{zubovas2020}. Instead, the jet could energetically fully sustain the outflows in these galaxies, as its power exceeds the outflow kinetic energy. To deduce the jet power, we used the observed $1.4\GHz$ flux of each galaxy as follows. The radio emission of the starburst (from its SNae) was computed from its far-infrared dust emission \citep{ivison2010} and subtracted from the observed $1.4\GHz$ flux. The residual synchrotron luminosity was converted into a total jet power, in other words the combined mechanical power of the cavity dominated by hadrons and radiative power of the synchrotron dominated by leptons, following a widely used calibration \citep{birzan2008}, which encapsulates that the jet power of the jet is typically dominated by the cavity power\citep{Cavagnolo2010}.

\begin{table*}[!ht]
    \caption{\label{table:literature}Outflow and AGN properties of galaxies with molecular outflows}
    \centering
    \begin{tabular}{|l l l l l l l|}
    \hline
        Galaxy & z & $V_\mathrm{out}$ & $\dot{M}_\mathrm{out}$ & $\log(\mathrm{L_{AGN}})$ & $F_\mathrm{1.4GHz}$ & $\log(P_\mathrm{jet})$ \\ 
        ~ & ~ & $\mathrm{km\,s^{-1}}$ & $\mathrm{M_\odot\,yr^{-1}}$& $\mathrm{erg\,s^{-1}}$ & $\mathrm{Jy}$ & $\mathrm{erg\,s^{-1}}$ \\ \hline
        NGC1377 & 0.00578 & 110 &  5 & 42.93 & <0.0015 \tablefootmark{1} & <42.46 \\ 
        NGC1068 & 0.00379 & 150  & 28 & 43.94 & 4.85 \tablefootmark{2} & 43.56 \\ 
        4C12.50 & 0.1217 & 800 \tablefootmark{3} & 652 \tablefootmark{3} & 45.7 & 5.1\tablefootmark{4} & 44.62 \tablefootmark{*} \\ 
        NGC1433 & 0.0036 & 100 & 0.7 & 42.24 & 0.0034 \tablefootmark{5} & 42.45 \\ 
        Mrk231 & 0.04217 & 700 & 350  & 45.72 & 0.31\tablefootmark{4} & ··· \\ 
        IC5063 & 0.011 & 300 & 10 & 44.3 & 1.87 \tablefootmark{2} & 43.74 \\ 
        M51 & 0.002 & 100 &  11 & 43.79 & 1.4 \tablefootmark{6} & 43.17 \tablefootmark{**} \\ 
        Mrk1014 & 0.16311 & 268  & 93 & 45.29 & 0.0225 \tablefootmark{1} & 43.84 \\ 
        NGC1266 & 0.00719 & 177 & 11 & 43.31 & 0.115 \tablefootmark{4} & 43.18 \\ 
        NGC5643 & 0.003999 & 500  & 85 & 42.91 & 0.203\tablefootmark{7} & 42.96 \\ 
        IRASF08572+3915 & 0.04311 & 800 & 403 & 45.72 & 0.0049 \tablefootmark{1} & ··· \\ 
        IRASF10565+2448 & 0.189 & 450 & 100 & 44.81 & 0.0576 \tablefootmark{1} & ··· \\ 
        IRASF11119+3257 & 0.06438 & 1000 & 203 & 46.2 & 0.106 \tablefootmark{8} & 43.77 \\ 
        IRASF23365+3604 & 0.02448 & 450  & 57 & 44.67 & 0.0287 \tablefootmark{4} & ··· \\ 
        NGC6240 & 0.123 & 400 & 267 & 45.38 & 0.427\tablefootmark{2} & ··· \\ 
        SDSSJ1356+1026 & 0.04256 & 500 & 118  & 46 & 0.49\tablefootmark{9} & ··· \\
        IRAS05189-2524 & 0.007268 & 491 & 219 & 44.47 & 0.0291 \tablefootmark{1} & ··· \\ 
        NGC4418 & 0.06764 & 134  & 19 & 43.81 & 0.0414\tablefootmark{4} & ··· \\ 
        IRASF14378-3651 & 0.030761 & 425 & 180 & 45.12 & 0.0349 \tablefootmark{1} & ··· \\ 
        IRAS13120-5453 & 0.08257 & 549 & 1115 & 44.35 & 0.065 \tablefootmark{2} & ··· \\ 
        IRASF14348-1447 & 0.04295 & 450 & 420 & 44.59 & 0.018 \tablefootmark{10}\tablefootmark{11} & ··· \\ 
        NGC613 & 0.00495 & 300 \tablefootmark{12} & 19 \tablefootmark{12} & 42.6 \tablefootmark{13} & 0.223 \tablefootmark{2} & 43.18 \\ 
        ESO420-613 & 0.01205 & 379 \tablefootmark{14} & 14 \tablefootmark{14} & 44.0 \tablefootmark{14} & 0.1 \tablefootmark{14} & 43.25 \\ 
        PKS1549-79 & 0.15 & 600 \tablefootmark{15} & 1733 \tablefootmark{15} & 44.7 \tablefootmark{25} & 5.36 \tablefootmark{16} & 44.69 \\
        3C273 & 0.15834 & 245 \tablefootmark{24} & 439 \tablefootmark{24} & 46.06 \tablefootmark{26} & 55 \tablefootmark{1} & 45.06 \tablefootmark{***} \\ 
        HE1353-1917 & 0.03502 & 145 \tablefootmark{17} & 22 \tablefootmark{17} & 43.42 \tablefootmark{18} & 0.0128 \tablefootmark{1} & ~ \\
        NGC6764 & 0.00807 & 170  & 1 & 42.23 & 0.113 \tablefootmark{19} & 43.15 \\
        Circinus & 0.0014 & 150 & 1 & 43.57 & 1.2 \tablefootmark{2} & 42.89 \\ 
        Mrk273 & 0.03777 & 620 & 200 & 44.16 & 0.145 \tablefootmark{1} & 43.62 \\ 
        NGC3256S & 0.00926 & 1400\tablefootmark{21} & 19 \tablefootmark{21}& 42.04 \tablefootmark{20} & 0.0168 \tablefootmark{****} & 42.97 \\ 
       NGC3100 & 0.0088 & 200 \tablefootmark{22} & 0.12 \tablefootmark{22} & 42.78 \tablefootmark{22} & 0.5 \tablefootmark{23} & 43 \tablefootmark{22} \\
       NGC6328 & 0.014 & 150 & 3.1 & 42.11 & 0.32 & 42.35  \\
       \hline
    \end{tabular}
    \tablefoot{are from \cite{allison2014}, unless otherwise indicated. For 4C12.50, properties are presented for the nuclear outflow component, as the extended components add little to the energetics \citep{fotopoulou2019}. Likewise, the infrared luminosity fraction of the southern nucleus comes from resolved SF rates \citep{lira2002}. The radio power, $P_\mathrm{jet}$, (for the sources with a certain radio jet detection) is calculated as in \cite{birzan2008} after the subtraction of the starburst contribution, as indicated in the text. Alternative radio power ($P_\mathrm{jet}$) computations are as follows: \tablefoottext{*}{43.96 for the synchrotron radiation \citep{fotopoulou2019}}, \tablefoottext{**}{39.3 for the X-ray cavity \citep{urquhart2018}}, and \tablefoottext{***}{an averaged value of 43.86 \cite{punsly2016}}. \tablefoottext{****}{To find the 1.4GHz flux fraction assigned to the southern nucleus of this galaxy merger, we used the information from spatially resolved 5 GHz data \citep{neff2003}}. \tablefoottext{1}{\cite{condon1998a}}. \tablefoottext{2}{\cite{allison2014}}. \tablefoottext{3}{\cite{fotopoulou2019}}. \tablefoottext{4}{\cite{baan2006}}. \tablefoottext{5}{\cite{ryder1996}}. \tablefoottext{6}{\cite{dumas2011}}.\tablefoottext{7}{\cite{condon1996}}. \tablefoottext{8}{\cite{urrutia2009a}}. \tablefoottext{9}{\cite{greene2012}}. \tablefoottext{10}{\cite{clemens2008}}. \tablefoottext{11}{\cite{clemens2010}}. \tablefoottext{12}{\cite{audibert2019}}. \tablefoottext{13}{\cite{davies2017}}. \tablefoottext{14}{\cite{fernandez-ontiveros2020}}.\tablefoottext{15}{\cite{oosterloo2019}}.\tablefoottext{16}{\cite{drake2004}}.\tablefoottext{17}{\cite{husemann2019}}.\tablefoottext{18}{\cite{shimizu2017}}.\tablefoottext{19}{\cite{marvil2015}}.\tablefoottext{20}{\cite{ohyama2015}}. \tablefoottext{21}{\cite{sakamoto2014a}}.\tablefoottext{22}{\cite{ruffa2022}}. \tablefoottext{23}{\cite{ruffa19a}}. \tablefoottext{24}{\cite{husemann2019a}}.\tablefoottext{25}{\cite{lanz2016}}.\tablefoottext{26}{\cite{gruppioni2016}}}
    
\end{table*}

\begin{figure*}[h]
    \begin{center}
        \includegraphics[width=17cm]{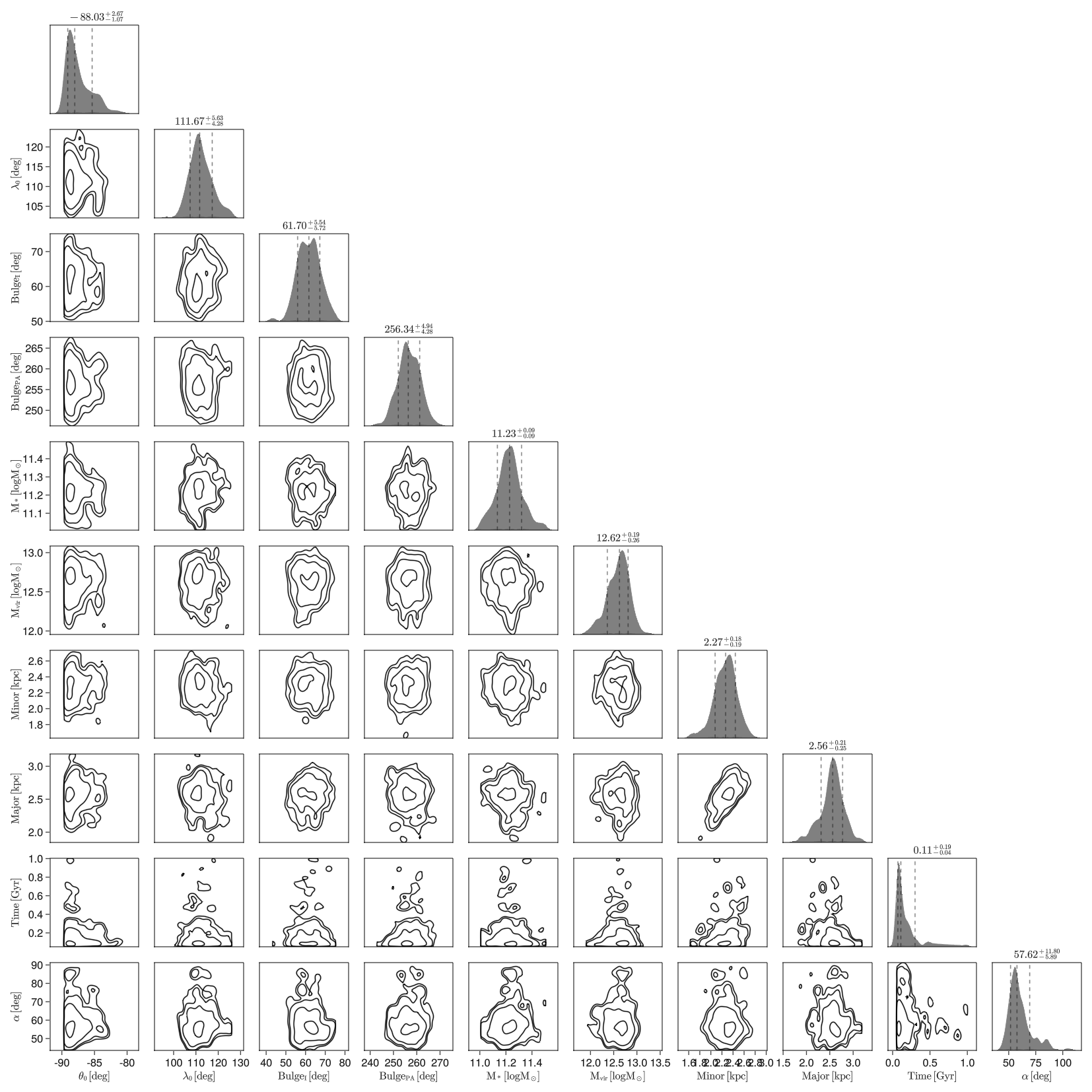}
        \caption{Corner plot of the parameters for the axisymmetric potential model.}
        \label{fig:corner_ax}
    \end{center}
\end{figure*}

\end{appendix}

\end{document}